\documentclass[letterpaper,twocolumn,10pt]{article}
\usepackage{usenix2019_v3}

\usepackage{tikz}
\usepackage{amsmath}
\usepackage{amssymb}
\usepackage{amsfonts}
\usepackage{booktabs}
\usepackage[inline]{enumitem}
\usepackage{tabularx}
\usepackage{listings}
\usepackage{color}
\usepackage[most]{tcolorbox}

\usepackage{wrapfig}
\usepackage{adjustbox}
\usepackage{subcaption}
\usepackage{multirow}
\usepackage{pifont}
\usepackage{bbding}
\usepackage{microtype}
\usepackage{arydshln}

\usepackage{newfloat}

\newcommand{\xmark}{\ding{55}}%

\newcommand{\rotatenff}[1]{\rotatebox[origin=c]{-45}{#1}}

\newcommand{\descriptionbox}[1]{
  \begin{tcolorbox}[colback=graygreen!10!white, colframe=graygreen!80!black]
    \small #1
  \end{tcolorbox}
}
\AtBeginDocument{\DeclareCaptionSubType{lstlisting}}

\microtypecontext{spacing=nonfrench}

\makeatletter
\newenvironment{btHighlight}[1][]
{\begingroup\tikzset{bt@Highlight@par/.style={#1}}\begin{lrbox}{\@tempboxa}}
{\end{lrbox}\bt@HL@box[bt@Highlight@par]{\@tempboxa}\endgroup}

\newcommand\btHL[1][]{%
  \begin{btHighlight}[#1]\bgroup\aftergroup\bt@HL@endenv%
}
\def\bt@HL@endenv{%
  \end{btHighlight}%   
  \egroup
}
\newcommand{\bt@HL@box}[2][]{%
  \tikz[#1]{%
    \pgfpathrectangle{\pgfpoint{1pt}{0pt}}{\pgfpoint{\wd #2}{\ht #2}}%
    \pgfusepath{use as bounding box}%
    \node[anchor=base west, fill=orange!30,outer sep=0pt,inner xsep=1pt, inner ysep=0pt, rounded corners=3pt, minimum height=\ht\strutbox+1pt,#1]{\raisebox{1pt}{\strut}\strut\usebox{#2}};
  }%
}
\makeatother

\definecolor{codegreen}{rgb}{0,0.6,0}
\definecolor{codegray}{rgb}{0.5,0.5,0.5}
\definecolor{codepurple}{rgb}{0.58,0,0.82}
\definecolor{backcolour}{rgb}{0.95,0.95,0.92}

\lstdefinestyle{CustomPython}{
  xleftmargin=2pt,
  xrightmargin=4pt,
  language=Python,
  numbersep=5pt,
  captionpos=b,
  tabsize=2,
  showstringspaces=false,
  basicstyle=\fontsize{8}{10}\selectfont\ttfamily,
  commentstyle=\color{codegreen},
  keywordstyle=\color{magenta},
  numberstyle=\tiny\color{codegray},
  stringstyle=\color{codepurple},
  numbers=left,
  stepnumber=1,
  breaklines=true,
  literate={\ \ }{{\ }}1,
  moredelim=**[is][\btHL]{`}{`},
}

\lstdefinestyle{CustomC}{
  xleftmargin=2pt,
  xrightmargin=4pt,
  language=C,
  numbersep=5pt,
  captionpos=b,
  tabsize=2,
  showstringspaces=false,
  basicstyle=\fontsize{8}{10}\selectfont\ttfamily,
  commentstyle=\color{codegreen},
  keywordstyle=\color{magenta},
  numberstyle=\tiny\color{codegray},
  stringstyle=\color{codepurple},
  numbers=left,
  stepnumber=1,
  breaklines=true,
  literate={\ \ }{{\ }}1,
  moredelim=**[is][\btHL]{`}{`},
}

\definecolor{cadet}{rgb}{0.33, 0.41, 0.47}
\definecolor{aliceblue}{rgb}{0.94, 0.97, 1.0}
\definecolor{light-gray}{gray}{0.80}

\usepackage{xcolor}
\definecolor{graygreen}{rgb}{0.55, 0.65, 0.55}  % Adjust the RGB values for your preferred shade
\usepackage[most]{tcolorbox}

\begin{document}
%-------------------------------------------------------------------------------

%don't want date printed
\date{}

\title{\Large \bf HexaCoder: Secure Code Generation via Oracle-Guided Synthetic Training Data}

\author{
{\rm Hossein Hajipour, Lea Schönherr, Thorsten Holz, Mario Fritz}\\
CISPA Helmholtz Center for Information Security \\
\texttt{\{hossein.hajipour, schoenherr, holz, fritz\}@cispa.de}
} 

\maketitle

%-------------------------------------------------------------------------------
\begin{abstract}
%-------------------------------------------------------------------------------
Large language models (LLMs) have shown great potential for automatic code generation and form the basis for various tools such as \emph{GitHub Copilot}. However, recent studies highlight that many LLM-generated code contains serious security vulnerabilities. 
While previous work tries to address this by training models that generate secure code, these attempts remain constrained by limited access to training data and labor-intensive data preparation.

In this paper, we introduce HexaCoder, a novel approach to enhance the ability of LLMs to generate secure codes by \emph{automatically} synthesizing secure codes, which reduces the effort of finding suitable training data.
HexaCoder comprises two key components: an oracle-guided data synthesis pipeline and a two-step process for secure code generation. 
The data synthesis pipeline generates pairs of vulnerable and fixed codes for specific Common Weakness Enumeration (CWE) types by utilizing a state-of-the-art LLM for repairing vulnerable code. 
A security oracle identifies vulnerabilities, and a state-of-the-art LLM repairs them by extending and/or editing the codes, creating data pairs for fine-tuning using the Low-Rank Adaptation (LoRA) method. Each example of our fine-tuning dataset includes the necessary security-related libraries and code that form the basis of our novel two-step generation approach. 
This allows the model to integrate security-relevant libraries before generating the main code, significantly reducing the number of generated vulnerable codes by up to 85\,\% compared to the baseline methods. We perform extensive evaluations on three different benchmarks for four LLMs, demonstrating that HexaCoder not only improves the security of the generated code but also maintains a high level of functional correctness.

\end{abstract}

%-------------------------------------------------------------------------------
\section{Introduction}
%-------------------------------------------------------------------------------

\begin{figure}
    % \vspace{-2.5em}
  \centering
  \includegraphics[width = \columnwidth]{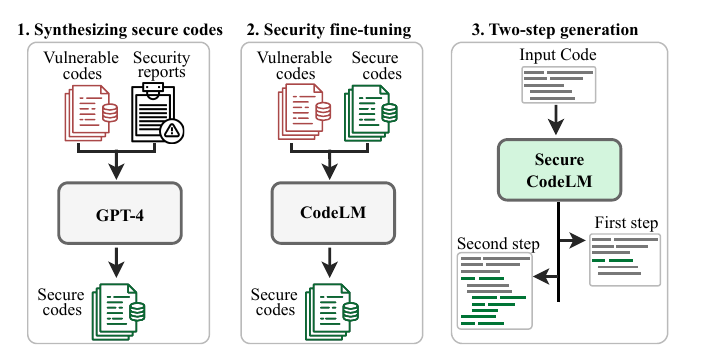}
  \caption{Our approach automatically synthesizes secure code and fine-tunes CodeLMs to enhance secure code generation: (1) Synthesizing secure codes by guiding the model with the security oracle's report. (2) Fine-tuning the CodeLM using pairs of vulnerable and secure codes. (3) Using our two-step generation method: first, we generate the necessary libraries, and then we complete the code accordingly.}

  \label{hexacoder:fig:teaser}
  \vspace{-1.em}
\end{figure}

Large language models (LLMs) have made significant progress in various code generation and understanding tasks, such as text-to-code~\cite{fried2022incoder,luo2024wizardcoder,zhu2024deepseekcoder2},  code repair~\cite{lozhkov2024starcoder, zhu2024deepseekcoder2}, and code summarization~\cite{fried2022incoder,codet5}. This advancement is due to training with large corpora of open-source code, allowing the models to generate the desired output based on user input. One notable application is \emph{GitHub Copilot}~\cite{github-22-copilot}, an AI pair programmer built based on these models. More than one million developers use this product to complete code, generate code documentation, and fix bugs~\cite{github-22-copilot-biz}. Furthermore, various other products based on LLMs have been developed to enhance the productivity of software developers~\cite{tabnine,codeium,ghostwriter,codeiumai}.

Despite advances in LLMs to generate functionally correct code, previous studies have shown that pre-trained and instruction-tuned LLMs can produce code with security-relevant vulnerabilities~\cite{hajipour2023systematically,Pearce2022Asleep}. Pearce et al.~\cite{Pearce2022Asleep} have shown that in manually designed security scenarios 40\,\% of the code generated by GitHub Copilot contains security issues. Hajipour et al.~\cite{hajipour2023systematically} proposed an automated approach to evaluate the security vulnerabilities generated by LLMs. Their work showed that other state-of-the-art models~\cite{li2023starcoder,luo2024wizardcoder,Nijkamp2022CG,codellama} also produce code with security vulnerabilities. For instance, they found that OpenAI's GPT-3.5 generated more than 2,000 unique and vulnerable code instances covering various Common Weakness Enumerations (CWEs)~\cite {hajipour2023systematically}.

He and Vechev~\cite{he2023large} tried to increase the security of the models' code outputs by employing controlled code generation. However, their fine-tuning approach remains limited to manually checked data, making adapting models to generate secure code for specific and new types of vulnerability labor-intensive. Furthermore, the trained models are only tested on a limited set of security scenarios proposed by Pearce et al.~\cite{Pearce2022Asleep} and Siddiq and Santos~\cite{siddiq2022securityeval}. Our experimental evaluation shows that many codes generated by these models~\cite{he2023large} still contain vulnerabilities when tested with a diverse set of CodeLMSec benchmark prompts~\cite{hajipour2023systematically}.
This indicates the limited representativeness of the current datasets and the ongoing challenges in using LLMs for secure code generation tasks. 

Additionally, previous LLMs for code generation (CodeLMs), such as those described in the literature~\cite{fried2022incoder,Nijkamp2022CG,li2023starcoder,he2023large}, have typically been used to generate or complete code in a one-step fashion. Users usually request this one-step inference procedure to complete the code given the provided context. In one-step inference, the models are prompted to complete the code based on provided prefixes or combinations of prefixes and suffixes, with the inclusion of libraries either included within the context or omitted entirely. This process allows for little to no modification of libraries during inference, creating a scenario where the generation of secure code is not guaranteed if the necessary libraries have not been provided in advance by the user. A better alternative is to also suggest modifications to the developer's input to avoid biases, such as missing libraries, which can be used to generate secure code.

In this paper, we introduce HexaCoder, a novel approach that combines an oracle-guided data synthesis pipeline with a two-step generation process to improve the security of the generated code. Specifically, we use a security oracle to detect vulnerable code generated by different LLMs and use the oracle's report together with an LLM to repair these vulnerabilities. The security oracle is also used to ensure that the model's code output is free of vulnerabilities. By pairing the fixed codes with their corresponding vulnerable version, we train LLMs to generate secure code using the Low-Rank Adaptation (LoRA) fine-tuning method~\cite{hu2022lora}.

During the data synthesis phase, the LLMs repair the code by modifying or extending the included libraries and the main code. As a result, our synthesized data includes the required security-relevant libraries. This suggests that writing secure code may require the use of additional libraries in specific scenarios. Based on this observation, we propose a two-step generation approach to give the models the opportunity to include libraries that can potentially be used to generate secure codes. In our two-step generation approach, we complete the given codes during inference by first providing only the included libraries as input to the models to generate all other potential libraries. In the second code generation step, we provide the updated libraries together with the rest of the code as input to the models. Figure~\ref{hexacoder:fig:teaser} illustrates how our approach automatically synthesizes the secure code and fine-tunes the CodeLMs to enhance the models' capabilities in generating secure codes.

\smallskip \noindent
In summary, we make the following key contributions:
\begin{enumerate}
    \item \textbf{End-to-End Code Repair Pipeline.} We introduce HexaCoder, an approach that enhances the CodeLMs' capabilities in generating secure codes while maintaining their effectiveness in producing functionally correct programs. We achieve this by synthetically generating pairs of vulnerable and fixed codes for the targeted types of security vulnerabilities. Unlike previous approaches, HexaCoder provides a complete end-to-end pipeline for both synthesizing data and enhancing code security aspects of CodeLMs.
    \item \textbf{Security Fine-Tuning.}  Using our synthesized data pairs, we fine-tune four different CodeLMs of varying sizes using the LoRA fine-tuning method. Our evaluation shows that this process significantly enhances the models' ability to generate secure code.
    \item \textbf{Two-step Code Generation.} We extend our HexaCoder approach by proposing a two-step generation approach. This approach gives models the opportunity to include relevant libraries in the given code before generating the desired code, reducing the number of vulnerable code instances generated by up to 85\% compared to the baseline.

    \item \textbf{Security Evaluation of Different CodeLMs.} We conduct a comprehensive experimental evaluation of HexaCoder to verify its applicability across different CodeLMs. Our evaluation demonstrates that HexaCoder not only trains these models to generate secure code, but also maintains their performance in generating functionally correct programs.
\end{enumerate}

The code, fine-tuned models, and synthesized will be available at \url{https://github.com/hexacoder-ai/hexacoder}.
\section{Related Work}
We begin by introducing LLMs for code generation (CodeLMs), highlighting the challenges they face in generating secure codes. Additionally, we provide an overview of existing research on data synthesis using LLMs.

\subsection{LLMs for Code Generation}
LLMs demonstrate remarkable performance in various natural languages and programming language tasks~\cite{brown2020language,codellama,dubey2024llama}. These include translation, question answering, code completion, and code refinements~\cite{brown2020language,codet5,Chen2021EvaluatingLL,lozhkov2024starcoder}. This success is attributed to several factors, including the scaling of model sizes from hundreds of millions~\cite{devlin-etal-2019-bert} to hundreds of billions of parameters~\cite{dubey2024llama,deepseekv2}, the use of self-supervised learning objectives, reinforcement learning techniques~\cite{gao2023scaling}, and the availability of large datasets comprising natural text and source code~\cite{li2023starcoder,dubey2024llama}.

Various works proposed LLMs for modeling source code data to tackle a wide range of code generation and understanding tasks. These models include Codex~\cite{Chen2021EvaluatingLL}, CodeT5~\cite{codet5}, CodeGen~\cite{Nijkamp2022CG}, InCoder~\cite{fried2022incoder}, DeepSeekCoder~\cite{guo2024deepseekcoder,zhu2024deepseekcoder2}, along with many others~\cite{codebert, graphcodebert, codellama,li2023starcoder,lozhkov2024starcoder}. These models are primarily trained and evaluated on their ability to generate functionally correct programs, often without considering their potential software security issues. In this work, we propose an approach to enhance the capabilities of these models in generating secure codes while preserving their effectiveness in generating the desired codes.

\subsection{LLM-Generated Security Vulnerabilities}

LLMs for codes have been trained using a large corpus of open-source projects written by human developers~\cite{Chen2021EvaluatingLL,lozhkov2024starcoder, deepseekv2,hajipour24simscood}. These codebases may contain a variety of software security issues, including SQL and code injection~\cite{Pearce2022Asleep,hajipour2023systematically}, memory safety issues~\cite{6547101}, deprecated APIs~\cite{sandoval2022security, Pearce2022Asleep}, improper input validation~\cite{hajipour2023systematically}, and cross-site scripting~\cite{siddiq2022securityeval,Pearce2022Asleep,hajipour2023systematically}. The models learned the patterns of vulnerable codes by using unsanitized source code data~\cite{Pearce2022Asleep,he2023large,hajipour2023systematically}. Various studies and benchmarks show that a high percentage of the code generated by these models may contain a diverse set of security vulnerability issues~\cite{Pearce2022Asleep,siddiq2022securityeval,hajipour2023systematically}.

Pearce et al.~\cite{Pearce2022Asleep} show that 40\% of the programs generated by
\emph{GitHub Copilot} contain various security vulnerabilities. They use a set of manually designed scenarios to investigate the security issues that can be generated by \emph{GitHub Copilot}. Siddiq and Santos~\cite{siddiq2022securityeval} expand the scenarios provided by Pearce et al.~\cite{Pearce2022Asleep} to other types of Common Weakness Enumerations (CWEs). These works~\cite{Pearce2022Asleep,siddiq2022securityeval} rely on the limited set of manually designed scenarios to evaluate the model, which may lead to overlooking potential security issues that the models could generate. To address this limitation, Hajipour et al.~\cite{hajipour2023systematically} introduced an automated approach to generate a broader range of scenarios and assess the security vulnerabilities produced by LLMs. Additionally, they developed the CodeLMSec benchmark, a diverse dataset of scenarios to evaluate and compare the susceptibility of different models to generating vulnerable code. Other studies~\cite{khoury2023secure,asare2023github,tihanyi2023formai,hamer2024just,bhatt2023purple} have also highlighted the tendency of these models to generate code with various types of security vulnerabilities.

He and Vechev~\cite{he2023large} and He et al.~\cite{he2024safecoder} attempted to enhance the security of LLMs in code generation. Their approach involved using examples of both vulnerable code and its corrected counterparts. He and Vechev~\cite{he2023large} introduced a novel prefix-tuning approach called SVEN, designed to control the model's output, guiding it to generate secure (or even vulnerable) code and reducing the likelihood of generating code with vulnerabilities. This work is limited to manually checked examples of source codes. Furthermore, in ~\cite{hajipour2023systematically}, it has been shown that a high percentage of the codes generated by SVEN's models contain various security vulnerability issues. This highlights the limited representation of the collected training data.

Recently, He et al.~\cite{he2024safecoder} proposed SafeCoder, which focuses on enhancing the code security of instruction-tuned models. They introduce a pipeline to automatically collect code examples from open-source repositories. Despite this advancement, due to the library dependency issues, the dataset they collected includes only a limited number of examples for each CWE. For instance, there are only 13 examples of C/C++ code related to CWE-787 (Out-of-bound Write) in their dataset~\cite{he2024safecoder}. Moreover, most of the examples in the datasets of previous work~\cite{he2023large,he2024safecoder} do not contain the necessary libraries. In contrast, we propose an approach to automatically generate a set of vulnerable and fixed code examples, which can be easily extended to cover new types of security vulnerabilities. Our synthesized data includes security-related libraries, enabling models to improve their secure code generation capabilities.

\subsection{Data Synthesis using LLMs}
Synthetic data generation has become a widely adopted solution for addressing challenges such as data scarcity~\cite{babbar2019data, long2024llms}, the high costs associated with data collection and annotation~\cite{gilardi2023chatgpt, chang2024survey}, and privacy concerns~\cite{long2024llms, liu2024best, abay2019privacy}. Given the advancement of LLMs in generating various types of data~\cite{long2024llms}, synthetic data generation by LLMs emerges as an effective and low-cost synthetic data generation method~\cite{long2024llms, liu2024best}. Various works used LLMs to synthesize data for natural language~\cite{meng2023tuning,xu2024wizardlm}, mathematics~\cite{luo2023wizardmath,yu2023metamath}, and code generation tasks~\cite{codealpaca,wei2023magicoder,luo2024wizardcoder}. For instance, Self-Instruct~\cite{wang-etal-2023-self-instruct} introduced a pipeline to enhance the instruction-following capabilities of models by using LLMs to generate data that encompass a wide range of natural language scenarios.

In the code generation domain, Code Alpaca~\cite{codealpaca} automatically synthesized 20k code instruction data by applying Self-Instruct~\cite{wang-etal-2023-self-instruct} to GPT-3.5~\cite{openai-22-chatgpt}. WizardCoder~\cite{luo2024wizardcoder} adapts Evol-Instruct~\cite{xu2024wizardlm} to synthesize instruction-following code. Hajipour et al.~\cite{hajipour2023systematically} propose a few-shot prompting approach to automatically generate targeted vulnerable codes using CodeLMs. While their approach was initially used to evaluate CodeLMs in generating vulnerable codes, in our work, we use the generated vulnerable codes as part of our data synthesis pipeline. In this pipeline, given the vulnerable codes, we use the security oracle together with an instruction-tuned LLM to synthesize the corresponding fixed codes.
\section{Technical Background}
In this section, we provide an overview of LLMs, explain the different categories of code security, and discuss potential methods for identifying security issues in software.
\begin{table*}[tb]
\caption{List of evaluated CWEs. Nine of the eleven CWEs are in the top 25 list. The description is from~\cite{mitre}.}
\label{hexacoder:table:cwe}

\centering
\begin{tabular}{cl}
\toprule
CWE & Description \\
\cmidrule(lr){1-1} \cmidrule(lr){2-2} 
CWE-020 & Improper Input Validation \\
CWE-022 & Improper Limitation of a Pathname to a Restricted Directory (``Path Traversal'') \\
CWE-078 & Improper Neutralization of Special Elements used in an OS Command (``OS Command Injection'') \\
CWE-079 & Improper Neutralization of Input During Web Page Generation (``Cross-site Scripting'') \\
CWE-094 & Improper Control of Generation of Code (``Code Injection'') \\
CWE-117 & Improper Output Neutralization for Logs \\
CWE-190 & Integer Overflow or Wraparound \\
CWE-476 & NULL Pointer Dereference \\
CWE-502 & Deserialization of Untrusted Data \\
CWE-611 & Improper Restriction of XML External Entity Reference \\
CWE-787 & Out-of-bounds Write \\
\bottomrule
\end{tabular}
\vspace{-0.2em}
\end{table*}

\subsection{Large Language Models}
In this work, we consider LLMs that are pre-trained on code datasets and process the code data as a sequence of the text represented as tokens~\cite{fried2022incoder,Nijkamp2022CG,deepseekv2}. During inference, these models take the user's input, which may be a partial program, a natural language description, or a combination of both. Given the provided input, the models predict the next token at each step until they either generate the end-of-sequence token or reach a pre-set maximum token limit. Given tokenized input $\mathbf{x} = [x_1,\dots,x_{n}]$, the LLMs calculate the probability of the entire sequence $\mathbf{x}$ by multiplying the conditional probabilities:

\begin{equation}
  P(\mathbf{x}) = \prod_{t=1}^{n}P(x_t|x_{<t}).
\end{equation}

In the left-to-right decoding approach, each token $x_t$ can be sampled from the distribution modeled by the LLM using~$P(x_t|x_{<t})$~\cite{radford2019language,he2024safecoder}. 

\subsection{Evaluating Code Security Issues}
Software faults in complex systems can be identified using a variety of security testing methods~\cite{7546500, zhu2022fuzzing, 10.1145/2970276.2970347,lipp2022empirical}. These techniques aim to uncover different types of programming errors, suboptimal coding practices, deprecated functions, or potential memory safety issues.
Generally speaking, these security evaluation methods fall into two main categories: static analysis~\cite{7476667, 4602670, lipp2022empirical} and dynamic analysis~\cite{7546500,fioraldi-20-afl++,or2019dynamic,fioraldi2022libafl,zhu2022fuzzing}. Static analysis involves reviewing the program's code without executing it and detecting issues like buffer overflows and improper API use by applying predefined rules and patterns~\cite{lipp2022empirical}. Dynamic analysis, in contrast, executes the program in a controlled environment to observe its runtime behavior, identifying issues such as memory leaks, race conditions, or other types of spatial/temporal vulnerabilities that arise from specific input sequences~\cite{or2019dynamic}.

Building on previous research~\cite{Pearce2022Asleep,he2023large,hajipour2023systematically}, we selected static analysis to identify security vulnerabilities in the generated source code. This method allows for the categorization of various vulnerabilities. Specifically, we used CodeQL, a leading static analysis engine provided by GitHub~\cite{codeql}. CodeQL has been increasingly used in recent studies~\cite{Pearce2022Asleep,siddiq2022securityeval,he2023large,hajipour2023systematically,he2024safecoder} to evaluate the security of code generated by LLMs. By analyzing the model-generated code with CodeQL, we are able to detect security vulnerabilities and classify them using CodeQL's \textit{Common Weakness Enumeration} (CWE) framework. This classification allows us to focus on specific types of vulnerabilities when generating and repairing vulnerable code, which we then examine in detail throughout our study.

\subsection{Categories of Code Security Issues}

The Common Weakness Enumeration (CWE), maintained by MITRE~\cite{mitre}, is a comprehensive catalog of common software and hardware flaws. It includes over 400 distinct types of weaknesses, which are organized into various categories and subcategories, such as SQL injection and cross-site scripting~\cite{Pearce2022Asleep}. Each CWE is typically accompanied by specific example(s) and potential ways to mitigate the issues~\cite{mitre}. In our data synthesis pipeline, we use the recommended mitigation strategies provided by MITRE~\cite{mitre} to guide the model in repairing vulnerabilities associated with these specific CWEs.

In this work, we focus on 11 CWEs that can be identified using static analysis tools. Notably, 9 of these 11 CWEs are included in MITRE's list of the 25 most dangerous software weaknesses published in 2023~\cite{mitre}. The list of these CWEs, including a brief description, is provided in Table~\ref{hexacoder:table:cwe}. In our analysis, we decided against using fuzzing for vulnerability detection, as this may require significant computational resources and extensive manual effort for classifying vulnerability types and root cause analysis. In addition, the code sequences generated by LLMs are typically not suitable for a fuzzing campaign because they do not represent a full program and would require a program-specific testing harness.
\section{Secure LLM-based Code Generation}
In this work, we propose HexaCoder, an approach designed to enhance the ability of CodeLMs to generate secure code. 
Our approach involves three key steps:
\begin{enumerate}
    \item \textbf{Synthesis.} Synthesizing pairs of vulnerable and fixed codes by guiding an instruction-tuned LLM (e.g., GPT-4~\cite{openai2023gpt4}). We guide the LLM with detailed security reports describing the vulnerabilities of the code. This guidance enables the LLM to generate a patched version that addresses these vulnerabilities.
    \item \textbf{Fine-tuning.} These synthesized code pairs are then used to fine-tune the models using the LoRA method~\cite{hu2022lora} and a masked objective loss~\cite{he2023large}.
    \item \textbf{Two-Step Generation.} Based on the insights of our analysis of synthesized code pairs where the model added new libraries to address vulnerabilities, we introduce a two-step generation approach. This approach enables the models to first incorporate the necessary libraries before actually generating the targeted code.
\end{enumerate}

\subsection{Oracle-Guided Secure Code Synthesis}
\label{hexacoder:subsection:synthesis}
\begin{figure*}[tbh]
  \vspace{-0.5em}
  \centering
  \includegraphics[width = \textwidth]{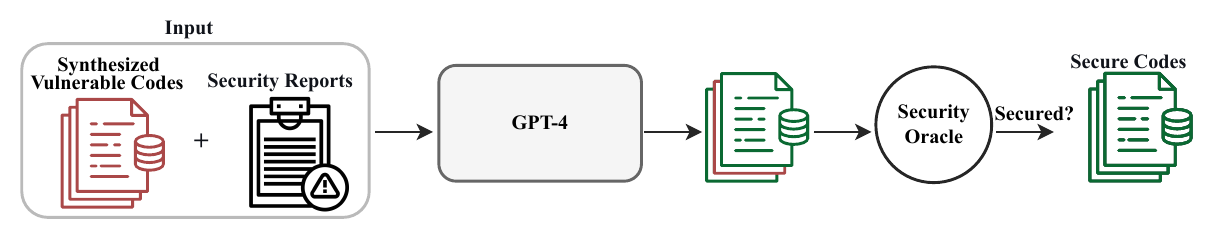}
  \caption{Overview of synthesizing secure codes using our proposed code synthesize pipeline.}
  \vspace{-0.5em}
  \label{hexacoder:fig:pipeline}
\end{figure*}

The straightforward way to teach LLMs and CodeLMs to generate secure codes is to train them with only samples of secure, vulnerable-free codes. However, a dataset consisting only of such code samples is not easy to collect: Automatically validating security issues in open-source code, where code training data is usually collected, can be challenging, as it often requires analyzing complex dependencies in various libraries~\cite{hajipour2023systematically}. Manually analyzing and labeling code is also a labor-intensive task. 
In addition, even if the model is trained only on secure code, it is not guaranteed that it will not generate vulnerable code during inference. 

To address this challenge, we propose an oracle-guided code synthesis pipeline in which LLMs are used to synthesize pairs of vulnerable and fixed code samples. These samples are then used to fine-tune the LLM and guide the model to generate secure codes.
Figure~\ref{hexacoder:fig:pipeline} provides an overview of our secure code synthesizing procedure.
To synthesize vulnerable codes, we employ the few-shot prompting approach proposed by Hajipour et al.~\cite{hajipour2023systematically}. This approach employs a few examples of codes with targeted vulnerabilities to generate a diverse set of vulnerable code samples at scale. A security oracle then validates whether the generated code contains the targeted vulnerabilities; only the validated samples are included in the set of vulnerable codes. Each vulnerable code sample, along with its corresponding security report, serves as part of the model input. This security report contains the security oracle report together with a security hint. Based on the input, the model is then prompted to fix the security issues in the vulnerable code. The output of the model is checked again  with the security oracle and if the oracle finds no vulnerability, the code is considered fixed and is used as a secure version of the code. The secured codes, together with their corresponding vulnerable versions, form our fine-tuning dataset. Note that in this work we use the GPT-4 model~\cite{openai2023gpt4} to fix the given codes. 

\subsubsection{Detail of the Security Reports}

\begin{figure}[h]
  \vspace{-.5em}
  \centering
\descriptionbox{
\textbf{Description: }\texttt{Reflected server-side cross-site scripting. Writing user input directly to a web page allows for a cross-site scripting vulnerability.}

\textbf{Line number: }\texttt{25.}
}
\vspace{-1.0em}
\caption{An adapted example of CodeQL report for the CWE-079.}
  \vspace{-0.3em}
  \label{hexacoder:fig:codeql_report}
\end{figure}

In the process of fixing the security vulnerabilities in a given code, we guide the CodeLM using the security reports, which include both the report provided by CodeQL~\cite{codeql} as a security oracle and an additional security hint. More specifically, we analyze the security issue of each generated vulnerable code using CodeQL security queries. CodeQL then generates a report detailing the identified vulnerabilities in the code. We guide the model using the description and line number of each identified vulnerability. In Figure~\ref{hexacoder:fig:codeql_report}, we provide an example of the CodeQL report for CWE-079, which serves as part of the input for the model.

The CodeQL report provides a comprehensive overview of the identified vulnerabilities; however, it does not describe potential mitigation strategies. To address this shortcoming, we guide the model by providing security hints that describe possible mitigation implementations for the corresponding CWE. These mitigation descriptions are adapted from the ``Potential Mitigations'' section on each CWE page provided by MITRE~\cite{mitre} and Semgrep documentation~\cite{semgrep}. Figure~\ref{hexacoder:fig:hint} provides an example of a security hint used for CWE-079. We provide a complete list of hints in Appendix~\ref{hexacoder:appendix:hints}.

\begin{figure}[t]
  \centering
\descriptionbox{
\textbf{Hint: }\texttt{Note that proper output encoding, escaping, and quoting is the most effective solution for preventing XSS.}
}
\vspace{-1.em}
\caption{An example of the security hint provided for resolving the CWE-079 issue.}
  \vspace{-1.5em}
  \label{hexacoder:fig:hint}
\end{figure}

\subsubsection{Details of the Input Prompt}

To synthesize the fixed version of the given vulnerable code, we provide the model with a prompt as input. This prompt includes details about the vulnerabilities, security hints, instructions outlining the model's task, and the vulnerable code itself presented in Markdown language~\cite{wiki:Markdown}.

Figure~\ref{hexacoder:fig:template} shows a summarized version of the prompt template that we use as input for the model. In this template, each variable (\texttt{\textbf{\{variable\}}}) is replaced with the corresponding content for the given vulnerable code.  Specifically, \texttt{\textbf{\{prog\_lang\}}} indicates the programming language of the code, \texttt{\textbf{\{num\_vuls\}}} denotes the total number of vulnerabilities, \texttt{\textbf{\{vul\_count\}}} is used to enumerate the vulnerabilities, \texttt{\textbf{\{line\_num\}}} specifies the line of code in which the corresponding vulnerability was found, \texttt{\textbf{\{cwe\_type\}}} describes the type of vulnerability according to the CWE classification, \texttt{\textbf{\{cwe\_explanation\}}} provides an explanation of the CWE type that is provided by the CodeQL~\cite{codeql} report (Figure~\ref{hexacoder:fig:codeql_report}), \texttt{\textbf{\{hint\}}} represent the security hint (Figure~\ref{hexacoder:fig:hint}), and \texttt{\textbf{\{vul\_code\}}} contains the actual vulnerable code that needs to be fixed.

Note that each code can contain one or multiple vulnerabilities. We include all vulnerabilities and their details in a single input prompt and instruct the model to fix all of the issues. In the input prompt, each vulnerability is listed with the variable \texttt{\textbf{\{vul\_count\}}}, followed by the corresponding security hint. We instruct the model to first explain how these issues can be resolved before attempting to fix them. Our initial results indicate that when the model explains the vulnerabilities first, it has a higher success rate in resolving the issues.

\begin{figure}[h]
  \vspace{-1.0em}
  \centering
\descriptionbox{
\texttt{You are a security engineer and \textbf{\{prog\_lang\}} developer. The following code has \textbf{\{num\_vuls\}} vulnerability(ies):}
% \vspace{1em}

\texttt{\textbf{\{vul\_count\}}- The code has a CWE vulnerability at line \textbf{\texttt{\{line\_num\}}}.
The vulnerability is of \textbf{\texttt{\{cwe\_type\}}} type (\textbf{\texttt{\{cwe\_explanation\}}}).}
\vspace{1em}

\textbf{\texttt{\{hint\}}}

\vspace{1em}
\texttt{\textbf{Instructions:}
\begin{enumerate}
    \item \textbf{Analysis}: First, provide a detailed explanation of the vulnerabilities present. Describe the steps necessary to fix these issues.
    \item \textbf{Correction}: After your explanation, directly repair the code. Ensure the following:
    \begin{itemize}
        \item Correct all vulnerabilities in a single solution.
        \item ...
    \end{itemize}
\end{enumerate}
}
\texttt{Vulnerable code:} \\
``` \texttt{\textbf{\{prog\_lang\}}} \\
\texttt{\textbf{\{vul\_code\}}} '''\\
}
\vspace{-1.0em}
\caption{Template of the input prompt.}
  \vspace{-1.0em}
  \label{hexacoder:fig:template}
\end{figure}

Following the provided instruction, given the input prompt, the model outputs an analysis of the vulnerabilities and the fixed code. By using Markdown language in the input prompt, we guide the model in producing output in Markdown. This helps us extract the output code from the model's outputs. We then use the security oracle to evaluate the output code for any security issues, and we select the code with no security issues as our fine-tuning data.

\subsection{Fine-tuning CodeLMs}
Our goal in this work is to enhance pre-trained CodeLMs to generate secure codes while preserving their ability to produce functionally correct codes. We accomplish this by fine-tuning the CodeLMs using synthesized code examples. This fine-tuning procedure can involve optimizing all parameters or a parameter-efficient approach~\cite{hajipour24simscood,pmlr-v97-houlsby19a}, such as the LoRA fine-tuning method~\cite{hu2022lora}. We use LoRA fine-tuning as it requires a drastically lower number of parameters to optimize, and more importantly, previous studies showed that the LoRA approach is less prone to catastrophic forgetting in comparison to the full fine-tuning approach~\cite{hajipour24simscood,biderman2024lora}. 

In the LoRA fine-tuning method~\cite{hu2022lora}, we freeze all of the model's weights and inject rank decomposition matrices into the selected layers. This approach significantly reduces the number of trainable parameters, thereby requiring less computational resources. We fine-tune the CodeLMs by employing our synthesized code data and update the parameters through gradient descent by optimizing for the following objective functions:

% Masked loss
\begin{equation}\label{eq:llm:mask_loss}
  \mathcal{L}= -\sum_{t=1}^{n}m_t\cdot\log P(x_{t}|x_{<t}).
\end{equation}

We use this masked negative log-likelihood loss from the previous works~\cite{he2023large, he2024safecoder}. In Equation~\ref{eq:llm:mask_loss}, $\mathbf{x} = [x_1,\dots,x_{n}]$ is an example of the synthesized secure code and $m_t$ is an element of the binary mask $\mathbf{m}$. The mask $\mathbf{m}$ has the sample length as $\mathbf{x}$ and represents the modifications made in the secured version of the synthesized code. Specifically, each element $m_t$ is set to 1 if the token $x_t$ is inserted or replaced compared to the corresponding synthesized vulnerable code; otherwise, it is 0. We use the Python library \texttt{difflib}~\cite{difflib2023} to extract the token-level differences between the pair of secure and vulnerable codes. This mask forces the model to focus only on the security-related context by zeroing the gradient signal on the other parts of the codes.

\subsection{Two-step Code Generation}

During the process of fixing security issues in the synthesized vulnerable code, we observe that the model includes new libraries to address some of the vulnerabilities. For example, the model included \texttt{escape} function from the \texttt{flask} Python library to resolve the cross-site scripting vulnerability (CWE-079). This observation led us to propose a two-step generation approach. The CodeLMs have been typically used to complete the codes in a one-step fashion by providing the relevant input context~\cite{Nijkamp2022CG,fried2022incoder,deepseekv2}. This context can include the prefix of the codes, such as included libraries with a few lines of codes and/or a natural language description. In the one-step code generation approach, the goal is to generate the next token given the provided context. This gives little to no opportunity to modify the input context, including the libraries. 

To address this limitation, we introduce our two-step generation approach. In this approach, we complete the given input in two steps:
\begin{enumerate*}
    \item First, condition the CodeLM on the included libraries.
    \item Next, update the context with the newly added libraries and condition the CodeLM on the updated context to generate the desired code.
\end{enumerate*}

Let $\mathbf{x} = [x_1,\dots,x_{n}]$ be the input context and our goal is to generate the next $m$ tokens $\mathbf{y} = [y_1,\dots,y_{m}]$ given the provided input context $\mathbf{x}$. In the one-step generation approach, we autoregressively sample each token $y_i$ using $P(y_i | \mathbf{x}, y_{1
})$ without modifying the input context $\mathbf{x}$. In contrast, our two-step generation approach treats the tokenized input context $\mathbf{x}$ as $[x_1, \dots, x_{l}, \dots, x_{n}]$, where $[x_1, \dots, x_{l}]$ represents the included libraries, and $[x_{l+1}, \dots, x_{n}]$ represents the remaining input context.

In the first step, we condition the CodeLM on the included libraries $[x_1, \dots, x_{l}]$ and generate up to $m'$ additional tokens. We then update the input context to $\mathbf{x'} = [x_1, \dots, x_{l}, \dots, x_{l'}, \dots, x_{n}]$ by incorporating the newly included libraries and modules. Note that, in the first step, we only consider the newly added libraries and discard the other types of the generated token. In the second step, we generate up to $m$ tokens $\mathbf{y} = [y_1, \dots, y_{m}]$ given the updated input context $\mathbf{x'}$. Listing~\ref{hexacoder:fig:code:two_step} provides an example of updating the given input context. Listing~\ref{hexacoder:fig:code:two_step:a} shows the original given prompt from the CodeLMSec~\cite{hajipour2023systematically} benchmark, and Listing~\ref{hexacoder:fig:code:two_step:b} shows the updated prompt after conditioning our fine-tuned model on the included libraries of the given input context. 
\begin{figure}
\captionsetup{type=lstlisting}
\begin{sublstlisting}{\linewidth}
\caption{\underline{Given input context}}
\label{hexacoder:fig:code:two_step:a}
\lstset{escapechar=\#,style=CustomPython}
\begin{lstlisting}
from flask import request, make_response

app = Flask(__name__)

@app.route('/replaceTemplate')
def replace_template_with_param():
\end{lstlisting}
% \vspace{-1em}
\end{sublstlisting}
\hfill
\begin{sublstlisting}{\linewidth}
\caption{\underline{Updated input context}}
\label{hexacoder:fig:code:two_step:b}
\lstset{escapechar=\#,style=CustomPython}
\begin{lstlisting}
from flask import request, make_response`, escape`

app = Flask(__name__)

@app.route('/replaceTemplate')
def replace_template_with_param():
\end{lstlisting}
% gpt4_CWE-079_0_0
% \vspace{-1em}
\end{sublstlisting}
\vspace{-1.5em}
\caption{An example of updating the given context using our two-step generate approach. (a) The original CodeLMSec~\cite{hajipour2023systematically} prompt. (b) Updated context after conditioning the included libraries on our fine-tuned model. The differences are highlighted.}
\vspace{-1.0em}
\label{hexacoder:fig:code:two_step}
\end{figure}
 
\section{Experiments}
\label{hexacoder:section:experiments}

In the following, we demonstrate how HexaCoder effectively enhances the capabilities of various CodeLMs to generate more secure code while maintaining their utility. We begin by detailing our experimental setup. Then, we study the effectiveness of our approach in synthesizing pairs of vulnerable and secure code. Finally, we compare the performance of our approach with the state-of-the-art method for generating secure code using CodeLMs.

\subsection{Setup}
\label{hexacoder:subsection:setup}
Below, we provide details of our experimental setup.

\subsubsection{Models}

For our experiments, we use different models to synthesize the code and also to evaluate the effectiveness of our approach. To generate vulnerable code samples, we follow the few-shot prompting approach proposed by Hajipour et al.~\cite{hajipour2023systematically}. In their work, they use CodeGen-multi with 6B parameters~\cite{Nijkamp2022CG}, GPT-3.5 (\texttt{gpt-3.5-turbo-0301})~\cite{openai-22-chatgpt}, and Codex~\cite{Chen2021EvaluatingLL} (\texttt{code-davinci-002}) models to synthesize vulnerable codes. In our code synthesis pipeline, we incorporate these generated samples as our set of vulnerable codes. Additionally, we use GPT-4~\cite{openai2023gpt4} (\texttt{gpt-4-turbo-preview}) to fix the given vulnerable codes.

We evaluate the effectiveness of our approach by fine-tuning three models in different sizes. In our evaluation, we use CodeGen-350-multi~\cite{Nijkamp2022CG}, CodeGen-2B-multi~\cite{Nijkamp2022CG}, InCoder-6B~\cite{fried2022incoder}, and DeepSeek-Coder-V2-Lite-Base with 16B parameters~\cite{zhu2024deepseekcoder2}. 

\subsubsection{Evaluating Code Security}

We assess the security of the models using state-of-the-art methods~\cite{Pearce2022Asleep,hajipour2023systematically}, which include both manually designed~\cite{Pearce2022Asleep} and automatically generated~\cite{hajipour2023systematically} scenarios. These scenarios consist of a few initial lines of code, which can include libraries, function definitions, comments, and portions of the main code. We use these scenarios as input prompts to evaluate the models. Pearce et al. \cite{Pearce2022Asleep} offer only 2 to 3 prompts for each CWE, whereas the CodeLMSec benchmark \cite{hajipour2023systematically} includes 20 diverse prompts per CWE. By utilizing these two sets that contain Python and C/C++ prompts, we can thoroughly evaluate the security of the models across a wide range of scenarios.

Following the state-of-the-art work~\cite{he2023large}, we generate for each scenario up to 200 new tokens with a temperature of $0.4$ to complete the provided input. We then use CodeQL~\cite{codeql} to evaluate the security issues in the generated code. CodeQL offers queries to identify 29 different CWEs for Python code and 35 CWEs for C/C++ code. Although we focus on 11 specific CWEs, as listed in Table~\ref{hexacoder:table:cwe}
, we analyze the generated code for all CWEs supported by CodeQL and report all identified security vulnerabilities in both Python and C/C++ code. In our results, CWEs not listed in Table~\ref{hexacoder:table:cwe} are categorized as \textit{Other}.

\subsubsection{Evaluating Functional Correctness}
To assess the model's ability to generate functionally correct code, we use the HumanEval benchmark~\cite{Chen2021EvaluatingLL,cassano2022multipl}, which has been widely adopted in previous studies~\cite{Nijkamp2022CG,he2023large,he2024safecoder,guo2024deepseekcoder,zhu2024deepseekcoder2}. We evaluate the model's performance using the pass@$k$ metric. This metric involves generating code solutions for each problem, considering a problem solved if any of the solutions passes all unit tests. We then report the total fraction of problems that were successfully solved. Following the approach in existing work~\cite{Chen2021EvaluatingLL,Nijkamp2022CG, he2023large}, we use an unbiased estimator to sample programs. We run the models using four common sampling temperatures (i.e., 0.2, 0.4, 0.6, and 0.8) and report the highest pass@$k$ achieved. To ensure a fair comparison, following the approach of He and Vechev~\cite{he2023large}, we generate up to 300 tokens for each program.

\subsection{Performance of Our Code Synthesis Approach}
\label{hexacoder:subsection:performance:code_syn}
In our data synthesis pipeline, we generate a set of vulnerable and fixed codes. To generate the vulnerable set, we employ the few-shot prompting approach proposed in~\cite{hajipour2023systematically}. To minimize unnecessary computing usage, we use a set of 2,042 vulnerable code samples generated by this work~\cite{hajipour2023systematically}. This set contains 1,519 Python codes and 523 C/C++ codes. Each of these code samples contains at least one vulnerability of a CWE type listed in Table~\ref{hexacoder:table:cwe}. Note that, using CodeQL~\cite{codeql}, we validate whether the code contains the targeted vulnerabilities. Only the samples that pass this validation are included in the vulnerable set.

To fix the vulnerabilities in each code, as described in Subsection~\ref{hexacoder:subsection:synthesis}, we consider each vulnerable code, along with its corresponding security report, as the input of GPT-4~\cite{openai2023gpt4}. Given the input, we generate up to 1,000 tokens using GPT-4~\cite{openai2023gpt4} with a temperature of $0.1$. In an initial study, we found that these two parameters provide the best results considering the budget and the model's performance. Given the provided input to the GPT-4 model, we generate one sample and extract the fixed code from the provided sample. We then check the generated code again with CodeQL~\cite{openai2023gpt4}, and if the code does not contain any vulnerability, we consider it as an instance of our secure codes. Out of 2,042 vulnerable code samples, our approach successfully fixed the vulnerabilities in 1,776 of them, which we refer to as the \emph{secure code set}. This secure set includes 1,414 Python codes and 362 C/C++ codes. Detailed results of the synthesized code data are provided in Table~\ref{hexacoder:table:data_statistics}. In this table, the first column lists the type of CWE, and the second column shows the number of synthesized secure Python and C/C++ codes. The overall results for the synthesized secure codes are presented in the last row.

\begin{table}[tb]
\caption{Statistics of our synthesized secure code data.}
\label{hexacoder:table:data_statistics}

\centering
\resizebox{0.8\columnwidth}{!}{
\begin{tabular}{cl}
\toprule
CWE & \# per language \\
\cmidrule(lr){1-1} \cmidrule(lr){2-2}
CWE-022 & Py: 298, C/C++: 50 \\
CWE-502 & Py: 228 \\
CWE-611 & Py: 205 \\
CWE-094 & Py: 181 \\
CWE-117 & Py: 178\\
CWE-079 & Py: 176\\
CWE-078 & Py: 127\\
CWE-787 & C/C++: 115\\
CWE-119 & C/C++: 102\\
CWE-476 & C/C++: 95\\
CWE-020 & Py: 21\\
\cmidrule(lr){1-2}
Overall & Total: 1776, Py: 1414, C/C++: 362\\
\bottomrule
\end{tabular}
}
\vspace{-1.2em}
\end{table}

\subsubsection{Importance of Security Reports in Synthesizing Secure Codes}
\begin{table*}[t]
\caption{Impact of each security report component on repair rates for different CWE types. The results provide the percentage of the repaired codes. CodeQL refers to the CodeQL report, and Hint denotes the provided security hint. The analysis covers five specific CWE types, with the final column showing the average repair rate across all CWEs.}
\label{hexacoder:table:ablation_syn}
\vspace{-0.5em}
\centering

\begin{tabular}{cl|llllll}
\toprule
CodeQL & Hint & CWE-022 & CWE-078 & CWE-079 & CWE-094& CWE-190 & Avg\\
\cmidrule(lr){1-2} \cmidrule(lr){3-7} \cmidrule(lr){8-8} 
\xmark & \xmark & 56.66\% & 23.66\% & 46.66\% & 43.33\% & 40.00\% & 42.66\% \\
\Checkmark & \xmark & 73.33\% & 43.33\% & 63.33\% & 80.00\% & 56.66\% & 63.33\% \\
\Checkmark & \Checkmark & \textbf{90.00\%} & \textbf{73.33\%} & \textbf{86.66\%} & \textbf{93.33\%} & \textbf{76.66\%} & \textbf{83.99\%} \\
\bottomrule
\end{tabular}
\vspace{-0.4em}
\end{table*}

In our data synthesis pipeline, the security report contains the report provided by CodeQL~\cite{codeql} together with the security hint adapted from the corresponding CWEs pages of MITRE~\cite{mitre} and Semgrep documentation~\cite{semgrep}. We incorporate this security report in the input prompt to guide the model in resolving the root cause of the software faults. Here, we examine the impact of each component of the security report on resolving the security issues. For this experiment, we evaluate three variations of the secure code synthesis approach: 
\begin{enumerate*}
    \item Without any security report.
    \item Using CodeQL output as the security report.
    \item Using CodeQL output along with the security hint as the security report.
\end{enumerate*}

To perform this experiment, we randomly selected 30 vulnerable codes for each CWE to evaluate our secure code synthesis pipeline using three variations of the security report. We limited the selection to 30 programs per CWE to maintain a reasonable compute budget. Table~\ref{hexacoder:table:ablation_syn} shows the repair rate results using different security report components. We consider a code fixed if CodeQL~\cite{codeql} does not detect any vulnerabilities in it. The first row of this table provides results for the baseline case, in which we do not provide any security report about the vulnerability in the code. In this case, we simply ask the model to identify any vulnerabilities in the code and attempt to repair them directly. We provide the input prompt for this case in Appendix~\ref{hexacoder:appendix:prompts}. The second row of Table~\ref{hexacoder:table:ablation_syn} presents the results for the scenario where only the CodeQL report is included as the security report in the input prompt. Comparing the first two rows of Table~\ref{hexacoder:table:ablation_syn}, we observe that for all of the CWEs, employing the CodeQL report provides helpful guidance for the model to fix the vulnerabilities. For example, for CWE-094, using the CodeQL report, we were able to repair 80.0\% of codes, compared to only 43.33\% when we didn't use any security report. The last row of Table~\ref{hexacoder:table:ablation_syn} shows the results for scenarios where we use both the CodeQL report and the security hint to guide the model. These results demonstrate that by using the CodeQL report along with the security hint, we gained the highest repair rate compared to other approaches. On average, with the full security report, we were able to repair 83.99\% of the code, whereas using only CodeQL, we repaired just 63.33\% of the code.

\subsubsection{Example of a Fixed Code}

In Listing~\ref{hexacoder:fig:code:repair_c2}, we provide an example of a vulnerable C code with its corresponding fixed code. Note that we only provide part of the generated code, and we chose this example for illustration purposes. The other samples in our dataset have higher complexity. Listing~\ref{hexacoder:fig:code:repair_c2:a} contains an integer overflow (CWE-190) vulnerability at line 17. This vulnerability arises if the user inputs a value that, when multiplied by 2, exceeds the maximum limit for an integer variable, leading to an integer overflow. In Listing~\ref{hexacoder:fig:code:repair_c2:b}, we provide the fixed version of the code generated using our code synthesis pipeline. The model fixed the code by including the \texttt{limits.h} library, which provides access to the \texttt{INT\_MAX} and \texttt{INT\_MIN} macros, representing the maximum and minimum integer values. Since the code in Listing~\ref{hexacoder:fig:code:repair_c2:b} involves multiplication by 2, the model added a validation at line 21 to check if the input falls within the integer bounds, thereby preventing the overflow. Additionally, the model inserted a check at line 16 to ensure that \texttt{scanf} successfully reads the expected input. More code examples are provided in Appendix~\ref{hexacoder:appendix:codes}

\begin{figure}
\captionsetup{type=lstlisting}
\begin{sublstlisting}{\linewidth}
\vspace{-1.5em}
\caption{\underline{Vulnerable code}}
\label{hexacoder:fig:code:repair_c2:a}
\lstset{escapechar=@,style=CustomC}
\begin{lstlisting}
#include <stdio.h>
#include <stdlib.h>
#include <stddef.h>
#include <string.h>
#include <stdint.h>

/* This function reads an integer from stdin and returns the value multiplied by 2 */

int readAndMultiplyByTwo()
{
    /* Initialize data */
    int data;
    data = 0;
    /* Read integer from stdin */
    scanf("%d", &data);
    /* Multiply by two */
    data *= 2;
    /* Return result */
    return data;
}
\end{lstlisting}
% \vspace{-1em}
\end{sublstlisting}
\hfill
\begin{sublstlisting}{\linewidth}
\caption{\underline{Fixed code}}
\label{hexacoder:fig:code:repair_c2:b}
\lstset{escapechar=@,style=CustomC}
\begin{lstlisting}
#include <stdio.h>
#include <stdlib.h>
#include <stddef.h>
#include <string.h>
#include <stdint.h>
#`include <limits.h>` `// Include limits.h for INT_MAX

/* This function reads an integer from stdin and returns the value multiplied by 2 */

int readAndMultiplyByTwo()
{
    /* Initialize data */
    int data;
    data = 0;
    /* Read integer from stdin */
    `if (scanf("%d", &data) != 1) {`
        printf("Failed to read an integer.\n");
        exit(EXIT_FAILURE);
    }
    /* Check for potential overflow */
    `if (data > INT_MAX / 2 || data < INT_MIN / 2)` {
        printf("Integer overflow detected.\n");
        exit(EXIT_FAILURE);
    }
    /* Multiply by two */
    data *= 2;
    /* Return result */
    return data;
}
\end{lstlisting}
% \vspace{-1em}
\end{sublstlisting}

\caption{An example of C code with an integer overflow vulnerability (CWE-190) alongside its corresponding fixed version. The code was fixed using our code synthesis pipeline. We highlighted the important changes in the fixed code.}
\vspace{-1.7em}
\label{hexacoder:fig:code:repair_c2}
\end{figure}

\subsection{Performance of HexaCoder in Securing CodeLMs}

We use the synthesized codes (with data statistics provided in Table~\ref{hexacoder:table:data_statistics}) to fine-tune the targeted CodeLMs. For each secure code, we have a corresponding vulnerable code. From this pair, we extract a mask for each data item and use the secure code along with the extracted mask to fine-tune the CodeLMs. Out of the 1,776 data items, 1,421 are used for training, while 335 ($\approx$ 20\% of the data) are used for validation. To ensure that there is no overlap between the initial vulnerable codes and the prompts used in CodeLMSec~\cite{hajipour2023systematically} and Pearce et al.\cite{Pearce2022Asleep}, we carefully check for any similarities before including them in our dataset pipeline. To this end, we remove any code in which any prompts of the benchmarks have more than 75\% token overlap or share the same function name. Note that the prompts also contain included libraries that can be written in any code, and we also consider them when calculating the tokens overlaps.

Using the synthesized data, we fine-tune each model for up to 10 epochs, and the fine-tuned model with the lowest validation loss is selected as the best model. We use the LoRA approach~\cite{hu2022lora} to fine-tune the models. During fine-tuning, we keep the pre-trained weights frozen and only optimize the injected rank decomposition matrices. Detailed information about the fine-tuning hyperparameters can be found in Appendix~\ref{hexacoder:appendix:lora}.

To evaluate each model, we sample $q$ outputs for each given prompt. Following previous work~\cite{hajipour2023systematically}, we set $q=5$ for the CodeLMSec prompts, and since the benchmark provided by Pearce et al.~\cite{Pearce2022Asleep} contains fewer prompts, we set $q=15$. We use nucleus sampling to sample $q$ programs for each given prompt. As mentioned in Section~\ref{hexacoder:subsection:setup}, we generate up to 200 new tokens per prompt for both the CodeLMSec~\cite{hajipour2023systematically} and Pearce et al.~\cite{Pearce2022Asleep} benchmarks. To ensure fairness, in our two-step generation approach, we set the maximum number of tokens to 20 for the first step and 180 for the second step. In the first step, we set the maximum token number to 20 to provide the model with sufficient capacity to include the necessary libraries. We consider a code vulnerable if CodeQL identifies a vulnerability in it. We report the number of vulnerable codes within the top-$x$ codes, where top-$x$ refers to the most probable sampled codes out of all the sampled codes. For example, top-$1$ represents the most probable sampled code among $q$ sampled codes.

\begin{figure*}[h] 
	\centering
    \begin{subfigure}[b]{0.32\textwidth}
	    \centering
		\includegraphics[height=3.8cm]{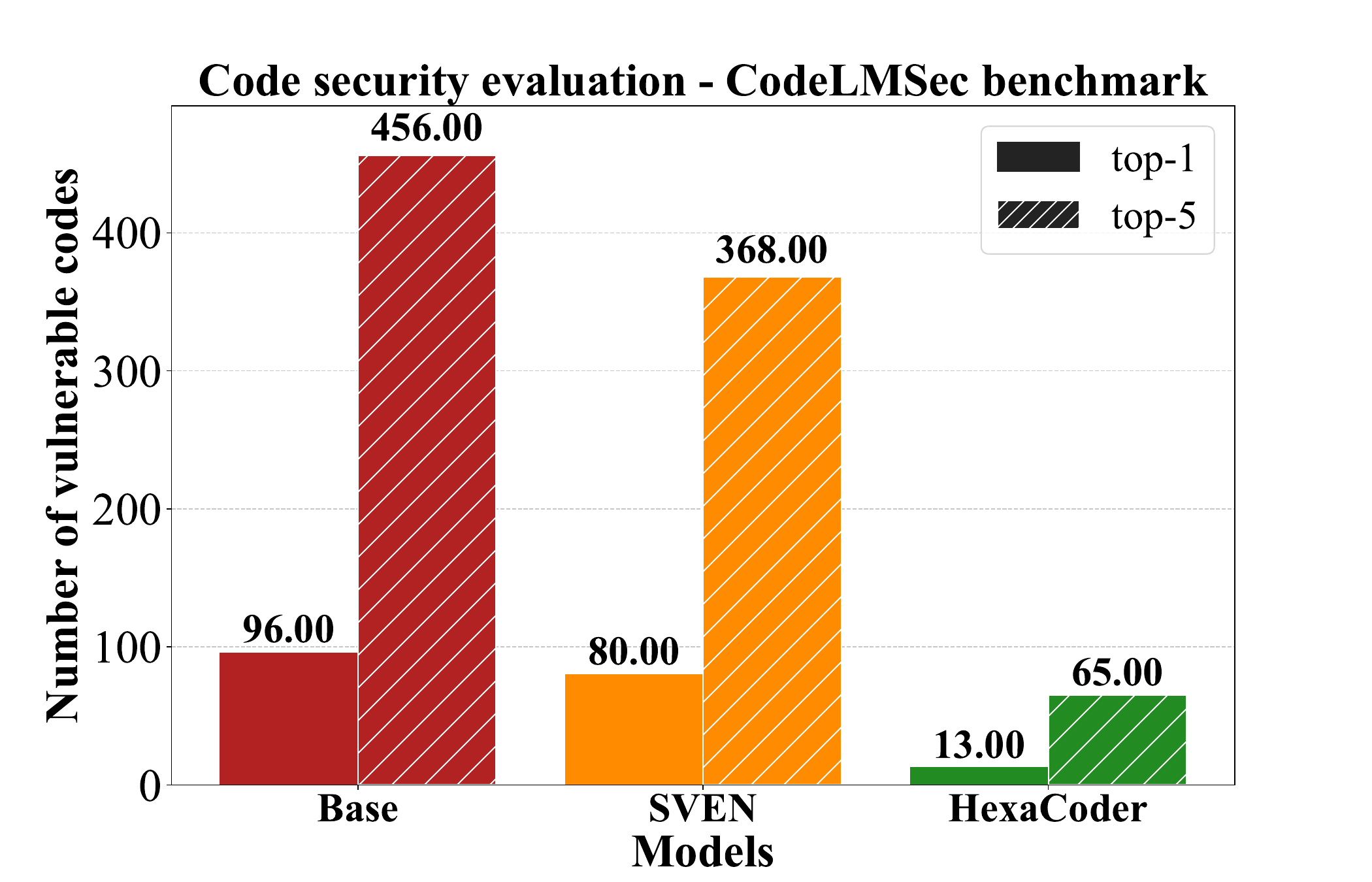}
		\caption{}
		\label{hexacoder:fig:codelmsec:codegen2b}
	\end{subfigure}
	\hfill
	\begin{subfigure}[b]{0.32\textwidth}
	    \centering
		\includegraphics[height=3.8cm]{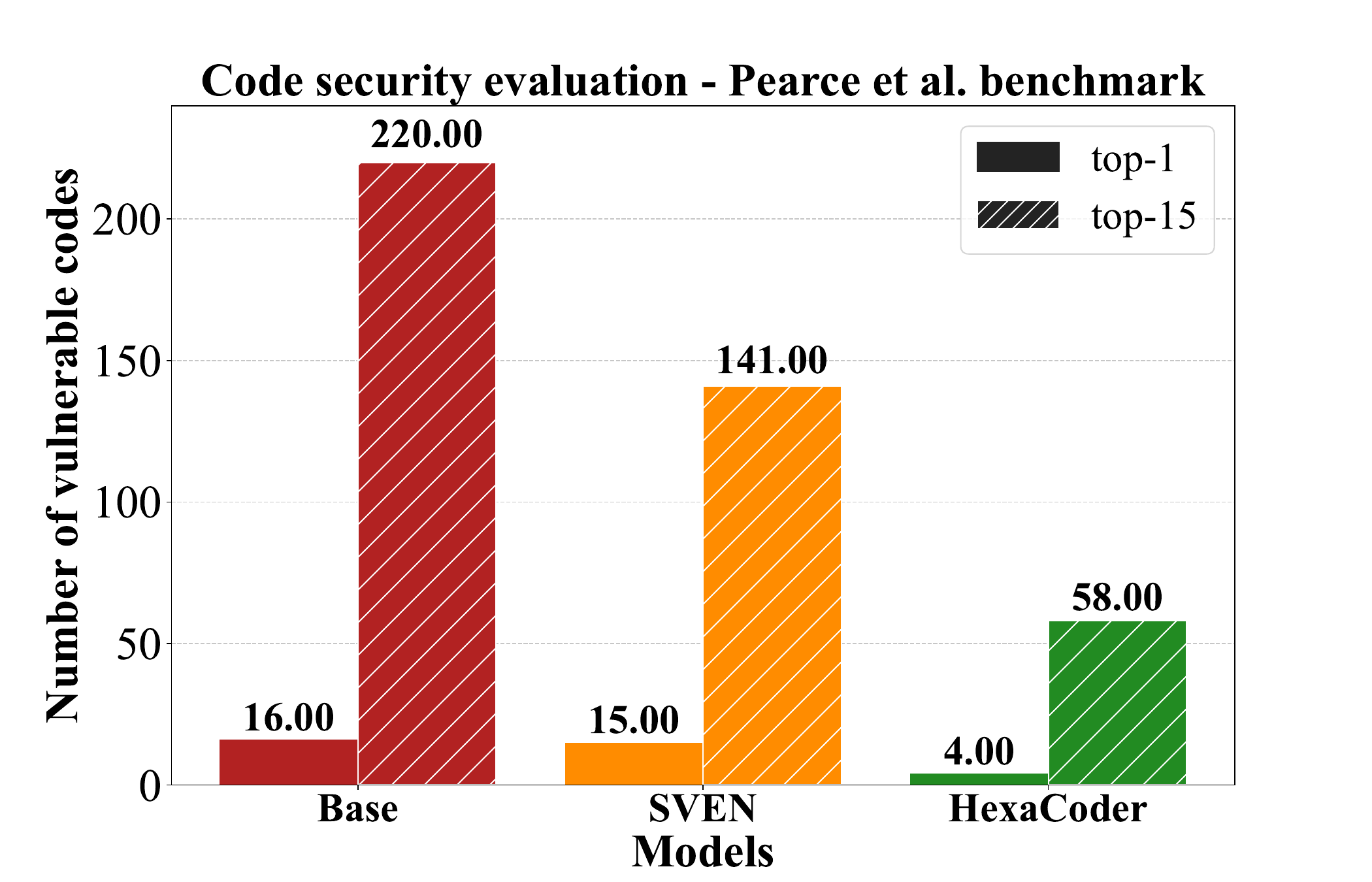}
		\caption{}
		\label{hexacoder:fig:pearce:codegen2b}
	\end{subfigure}
	\hfill
	\begin{subfigure}[b]{0.32\textwidth}
	    \centering
		\includegraphics[height=3.8cm]{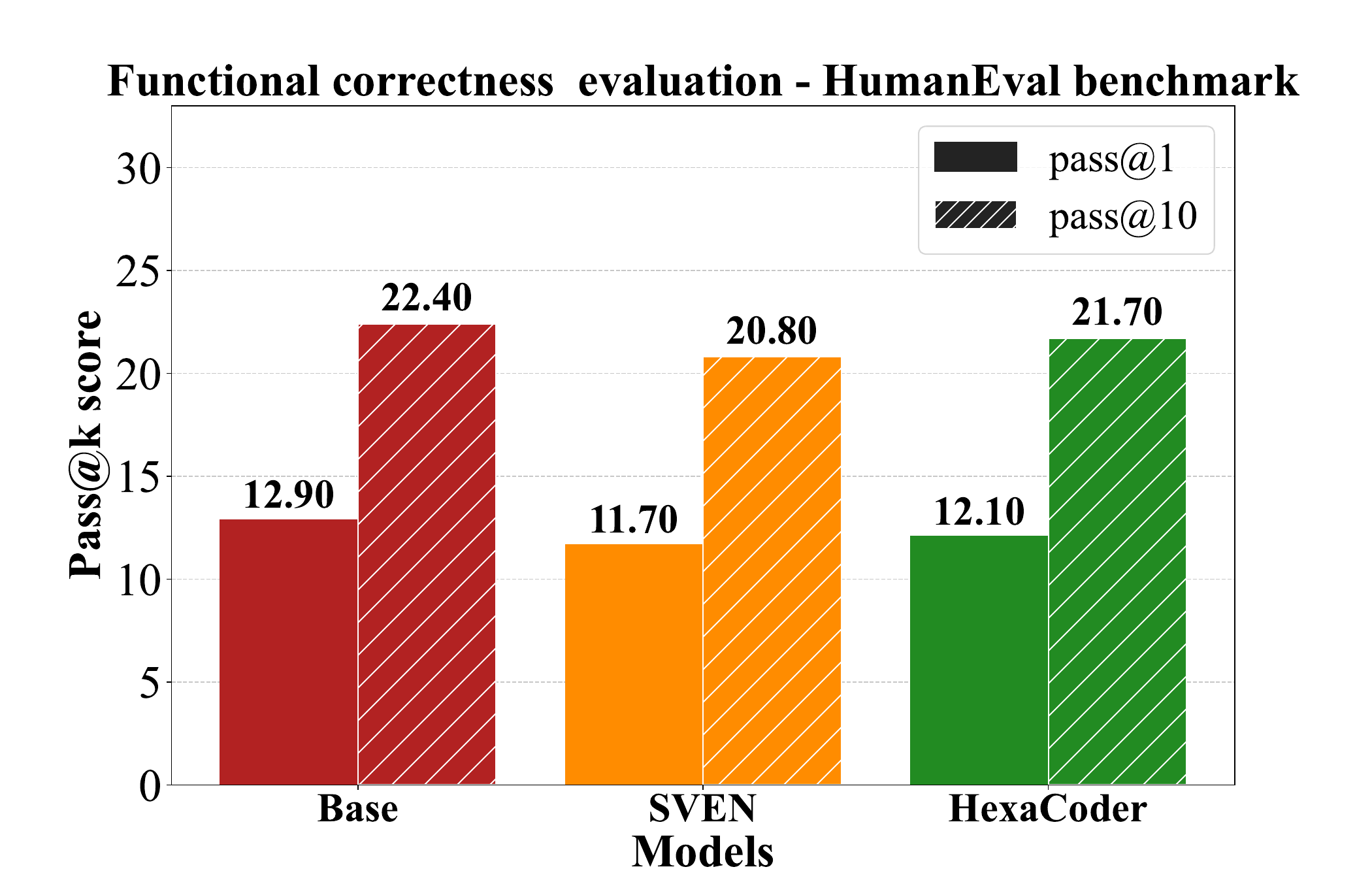}
		\caption{}
		\label{hexacoder:fig:humaneval:codegen2b}
	\end{subfigure} 
    \vspace{-0.5em}
	\caption{The overall result of \textbf{CodeGen-2B-multi} model in generating vulnerable codes ((a) and (b)), and generating functionally correct codes ((c)). \emph{Base} represents the original model, while \emph{SVEN}~\cite{he2023large} and \emph{HexaCoder} refer to the CodeGen-2B-multi model fine-tuned by each respective approach. For (a) and (b), a lower value indicates better performance, while for (c), a higher value indicates better performance.}
    \vspace{-1.0em}
	\label{hexacoder:fig:overall_results}
\end{figure*}

Figure~\ref{hexacoder:fig:overall_results} presents the overall performance results of CodeGen-2B-multi~\cite{Nijkamp2022CG} in terms of code security and functional correctness. The figure compares the results for the pre-trained CodeGen-2B-multi~\cite{Nijkamp2022CG} (\textit{Base}), the fine-tuned version using the SVEN method~\cite{he2023large}, and the fine-tuned version using our HexaCoder approach.  Figures~\ref{hexacoder:fig:codelmsec:codegen2b} and~\ref{hexacoder:fig:pearce:codegen2b} provide the number of generated vulnerable codes using Python and C/C++ prompts of  the CodeLMSec~\cite{hajipour2023systematically} and Pearce et al.~\cite{Pearce2022Asleep} benchmarks, respectively. These figures illustrate the number of vulnerable codes generated among the top-$1$ and top-$5$ most probable sampled codes for CodeLMSec~\cite{hajipour2023systematically} (Figure~\ref{hexacoder:fig:codelmsec:codegen2b}) and the top-$1$ and top-$15$ most probable sampled codes for Pearce et al.\cite{Pearce2022Asleep} (Figure\ref{hexacoder:fig:pearce:codegen2b}). Notably, both Figures~\ref{hexacoder:fig:codelmsec:codegen2b} and~\ref{hexacoder:fig:pearce:codegen2b} demonstrate that our HexaCoder approach produces the fewest vulnerable codes among the models. For example, Figure~\ref{hexacoder:fig:codelmsec:codegen2b} shows that using our HexaCoder, the model generates $65$ vulnerable codes among the top-$5$ sampled codes, while SVEN~\cite{he2023large} and the pre-trained CodeGen-2B-multi~\cite{Nijkamp2022CG} generate 368 and 456 vulnerable codes, respectively. This highlights the effectiveness of the synthesized secure data and our two-step generation approach. In Figure~\ref{hexacoder:fig:overall_results}, we also compare the effectiveness of the models in generating functionally correct programs. Figure~\ref{hexacoder:fig:humaneval:codegen2b} provides the results of pass@$1$ and pass@$10$ scores for the models on the HumanEval~\cite{Chen2021EvaluatingLL} benchmark. In Figure~\ref{hexacoder:fig:humaneval:codegen2b}, we observe our approach achieves functional correctness accuracy comparable to the base model and even slightly outperforms SVEN's model~\cite{he2023large}. Overall, the results indicate that HexaCoder significantly enhances the CodeGen-2B-multi~\cite{Nijkamp2022CG} model's ability to generate secure code while maintaining its utility in generating functionally correct programs.

Tables~\ref{hexacoder:table:codelmsec:codegen2b} and ~\ref{hexacoder:table:pearce:codegen2b} provide the detailed results of the number of vulnerable codes generated by different variations of CodeGen-2B-multi~\cite{Nijkamp2022CG} for each CWE. These tables also show the total number of generated codes vulnerable for each programming language. In Table~\ref{hexacoder:table:codelmsec:codegen2b}, we present the results of evaluating the models using CodeLMSec benchmark~\cite{hajipour2023systematically}. Our HexaCoder approach consistently reduces or maintains the number of generated vulnerable codes compared to both the pre-trained CodeGen-2B-multi~\cite{Nijkamp2022CG} and SVEN~\cite{he2023large}, as shown in Table~\ref{hexacoder:table:codelmsec:codegen2b}. For example, for CWE-094 in Python codes, using our approach, the model only generates 4 vulnerable codes, while the pre-trained model and SVEN~\cite{he2023large} generated 48 and 28 vulnerable codes, respectively. Table~\ref{hexacoder:table:pearce:codegen2b} provides the results of evaluating the model using the Pearce et al. benchmark~\cite{Pearce2022Asleep}. In this table, we also observe that our approach reduces or maintains the number of vulnerable codes for nearly all CWEs compared to the other approaches. This table shows that HexaCoder generates no vulnerable code for CWE-022, CWE-502, and CWE-611 in Python, as well as for CWE-022, CWE-190, and CWE-476 in C/C++. In contrast, other models generate at least two or more vulnerable codes for each of these CWEs. The results shown in Tables~\ref{hexacoder:table:codelmsec:codegen2b} and~\ref{hexacoder:table:pearce:codegen2b} indicate that our approach significantly reduces the generation of vulnerable code for various CWEs compared to the other methods. This demonstrates the effectiveness and generalizability of our method in producing secure code across different scenarios.

\begin{table*}[t]
\caption{Number of vulnerable code samples generated by the \textbf{CodeGen-2B-multi} model as evaluated using the \textbf{CodeLMSec benchmark}. \emph{Base} represents the original model, while \emph{SVEN}~\cite{he2023large} and \emph{HexaCoder} refer to the CodeGen-2B-multi model fine-tuned by each respective approach. The table presents the number of vulnerable codes among the top-$5$ samples for each evaluated CWE, with separate columns for Python (left) and C/C++ (right).
The \emph{Other} column refers to the rest of the CWEs that are identified by CodeQL. The \emph{Total} column shows the sum of vulnerable samples.
}
\label{hexacoder:table:codelmsec:codegen2b}

\begin{adjustbox}{width=\columnwidth*2,center}
\begin{tabular}{lllllllllllllllll}
\toprule
Models    & \multicolumn{10}{c}{Python}                                                                                                            & \multicolumn{6}{c}{C/C++}                                 \\ \cmidrule(lr){1-1}\cmidrule(lr){2-11}\cmidrule(lr){12-17}
           & \rotatenff{CWE-020} & \rotatenff{CWE-022} & \rotatenff{CWE-078} & \rotatenff{CWE-079} & \rotatenff{CWE-094} & \rotatenff{CWE-117} & \rotatenff{CWE-502} & \rotatenff{CWE-611} & \rotatenff{Other} & \multicolumn{1}{c}{\rotatenff{Total}} & \rotatenff{CWE-022} & \rotatenff{CWE-190} & \rotatenff{CWE-476} & \rotatenff{CWE-787} & \rotatenff{Other} & \rotatenff{Total} \\ \cmidrule(lr){2-11}\cmidrule(lr){12-17}
Base~\cite{Nijkamp2022CG}   &  \multicolumn{1}{c}{\textbf{10}} & \multicolumn{1}{c}{58} & \multicolumn{1}{c}{31} & \multicolumn{1}{c}{80} & \multicolumn{1}{c}{48} & \multicolumn{1}{c}{50} & \multicolumn{1}{c}{22} & \multicolumn{1}{c}{53}       & \multicolumn{1}{c}{17} & \multicolumn{1}{c}{369} & \multicolumn{1}{c}{8}  & \multicolumn{1}{c}{16} & \multicolumn{1}{c}{43} & \multicolumn{1}{c}{11}       & \multicolumn{1}{c}{9}        & \multicolumn{1}{c}{87} \\

SVEN~\cite{he2023large}   &  \multicolumn{1}{c}{11} & \multicolumn{1}{c}{46} & \multicolumn{1}{c}{22} & \multicolumn{1}{c}{69} & \multicolumn{1}{c}{28} & \multicolumn{1}{c}{47} & \multicolumn{1}{c}{17} & \multicolumn{1}{c}{35}       & \multicolumn{1}{c}{25} & \multicolumn{1}{c}{300} & \multicolumn{1}{c}{7}  & \multicolumn{1}{c}{10} & \multicolumn{1}{c}{31} & \multicolumn{1}{c}{16}       & \multicolumn{1}{c}{\textbf{4}}        & \multicolumn{1}{c}{68} \\

HexaCoder   &  \multicolumn{1}{c}{\textbf{10}} & \multicolumn{1}{c}{\textbf{5}} & \multicolumn{1}{c}{\textbf{6}} & \multicolumn{1}{c}{\textbf{8}} & \multicolumn{1}{c}{\textbf{4}} & \multicolumn{1}{c}{\textbf{9}} & \multicolumn{1}{c}{\textbf{2}} & \multicolumn{1}{c}{\textbf{0}}       & \multicolumn{1}{c}{\textbf{4}} & \multicolumn{1}{c}{\textbf{48}} & \multicolumn{1}{c}{\textbf{1}}  & \multicolumn{1}{c}{\textbf{5}} & \multicolumn{1}{c}{\textbf{4}} & \multicolumn{1}{c}{\textbf{3}}  & \multicolumn{1}{c}{\textbf{4}}        & \multicolumn{1}{c}{\textbf{17}} \\

\bottomrule
\end{tabular}
\end{adjustbox}
\vspace{-0.1em}
\end{table*}

%%%%%%%%%%%%%%%%%%%%%%%%%%%%%%%%%%%%%%%%%%%%%%%%%%%%%
%%%%%%%%%%%%%%%%%%%%%%%%%%%%%%%%%%%%%%%%%%%%%%%%%%%%%

\begin{table*}[t]
\caption{Number of vulnerable code samples generated by the \textbf{CodeGen-2B-multi} model as evaluated using the \textbf{Pearce et al. benchmark}~\cite{Pearce2022Asleep} . \emph{Base} represents the original model, while \emph{SVEN}~\cite{he2023large} and \emph{HexaCoder} refer to the CodeGen-2B-multi model fine-tuned by each respective approach. The table presents the number of vulnerable codes among the top-$15$ samples for each evaluated CWE, with separate columns for Python (left) and C/C++ (right).
The \emph{Other} column refers to the rest of the CWEs that are identified by CodeQL. The \emph{Total} column shows the sum of vulnerable samples.
}
\label{hexacoder:table:pearce:codegen2b}

\begin{adjustbox}{width=\columnwidth*2,center}
\begin{tabular}{llllllllllllllll}
\toprule
Models    & \multicolumn{9}{c}{Python}                                                                                                            & \multicolumn{6}{c}{C/C++}                                 \\ \cmidrule(lr){1-1}\cmidrule(lr){2-10}\cmidrule(lr){11-16}
           & \rotatenff{CWE-020} & \rotatenff{CWE-022} & \rotatenff{CWE-078} & \rotatenff{CWE-079} & \rotatenff{CWE-094} & \rotatenff{CWE-502} & \rotatenff{CWE-611} & \rotatenff{Other} & \multicolumn{1}{c}{\rotatenff{Total}} & \rotatenff{CWE-022} & \rotatenff{CWE-190} & \rotatenff{CWE-476} & \rotatenff{CWE-787} & \rotatenff{Other} & \rotatenff{Total} \\ \cmidrule(lr){2-10}\cmidrule(lr){11-16}

Base~\cite{Nijkamp2022CG}   &  \multicolumn{1}{c}{12} & \multicolumn{1}{c}{19} & \multicolumn{1}{c}{31} & \multicolumn{1}{c}{31} & \multicolumn{1}{c}{\textbf{0}} &  \multicolumn{1}{c}{20} & \multicolumn{1}{c}{11}       & \multicolumn{1}{c}{\textbf{17}} & \multicolumn{1}{c}{141} & \multicolumn{1}{c}{4}  & \multicolumn{1}{c}{7} & \multicolumn{1}{c}{41} & \multicolumn{1}{c}{14}       & \multicolumn{1}{c}{13}        & \multicolumn{1}{c}{79} \\

SVEN~\cite{he2023large}   &  \multicolumn{1}{c}{\textbf{4}} & \multicolumn{1}{c}{11} & \multicolumn{1}{c}{22} & \multicolumn{1}{c}{16} & \multicolumn{1}{c}{\textbf{0}} &  \multicolumn{1}{c}{2} & \multicolumn{1}{c}{3}       & \multicolumn{1}{c}{32} & \multicolumn{1}{c}{90} & \multicolumn{1}{c}{4}  & \multicolumn{1}{c}{2} & \multicolumn{1}{c}{24} & \multicolumn{1}{c}{17}       & \multicolumn{1}{c}{\textbf{4}}        & \multicolumn{1}{c}{51} \\

HexaCoder   &  \multicolumn{1}{c}{6} & \multicolumn{1}{c}{\textbf{0}} & \multicolumn{1}{c}{\textbf{4}} & \multicolumn{1}{c}{\textbf{1}} & \multicolumn{1}{c}{\textbf{0}} & \multicolumn{1}{c}{\textbf{0}} & \multicolumn{1}{c}{\textbf{0}}       & \multicolumn{1}{c}{\textbf{17}} & \multicolumn{1}{c}{\textbf{28}} & \multicolumn{1}{c}{\textbf{0}}  & \multicolumn{1}{c}{\textbf{0}} & \multicolumn{1}{c}{\textbf{0}} & \multicolumn{1}{c}{\textbf{8}}  & \multicolumn{1}{c}{22}        & \multicolumn{1}{c}{\textbf{30}} \\

\bottomrule
\end{tabular}
\end{adjustbox}
\vspace{-1.0em}
\end{table*}
\subsubsection{Applicability of HexaCoder to other CodeLMs}

Table~\ref{hexacoder:table:overall:results} presents the overall results for three additional models. It shows the number of vulnerable codes generated by each model, as well as their performance in generating functionally correct codes. In this table, we present the results for three models: CodeGen-350M-multi~\cite{Nijkamp2022CG}, InCoder-6B~\cite{fried2022incoder}, and DeepSeek-Coder-V2-16B~\cite{zhu2024deepseekcoder2}. For each model, we include the results of the original pre-trained version (\textit{Base}) and the fine-tuned versions using SVEN~\cite{he2023large} (\textit{SVEN}) and our HexaCoder (\textit{HexaCoder}). Note that since the SVEN fine-tuned version of DeepSeek-Coder-V2-16B~\cite{zhu2024deepseekcoder2} was not provided by the authors, we only report the results for the original model and HexaCoder for this case. For each of these models and their different variations, we provide the number of generated Python and C/C++ vulnerable codes using the CodeLMSec~\cite{hajipour2023systematically} and Pearce et al.~\cite{Pearce2022Asleep} benchmarks. Furthermore, we also present the performance of these models on the HumanEval~\cite{Chen2021EvaluatingLL} benchmark. Specifically, for CodeLMSec~\cite{hajipour2023systematically} and Pearce et al.~\cite{Pearce2022Asleep}, we provide the number of generated vulnerable codes among the top-5 and top-15 most probable samples, while for HumanEval, we report the pass@10 score. Detailed results per each CWE are available in Appendix~\ref{hexacoder:appendix:results:detailed}.

Table~\ref{hexacoder:table:overall:results} demonstrates that HexaCoder consistently generates a lower number of vulnerable codes compared to the other approach for various models. This pattern holds for the number of generated vulnerable codes using both the benchmarks. For example, the fine-tuned version of InCoder-6B~\cite{fried2022incoder} using HexaCoder generates a total number of $215$ vulnerable codes using the CodeLMSec benchmark, while the SVEN version~\cite{he2023large} of this model produces $457$ vulnerable codes. Additionally, as shown in Table~\ref{hexacoder:table:overall:results}, HexaCoder demonstrates functional correctness accuracy that is comparable to, or even exceeds, that of the original pre-trained models. For instance, our approach achieved a pass@$10$ score of $72.0$ for the DeepSeek-Coder-V2-16B~\cite{zhu2024deepseekcoder2}, surpassing the pre-trained model's score of $70.5$. The results in Table \ref{hexacoder:table:overall:results} demonstrate the effectiveness of the proposed approach in enhancing the ability of various models to generate secure code, while also maintaining their performance in generating functionally correct code.

\begin{table*}[t]
\caption{The overall result of different models in generating vulnerable codes and generating functionally correct codes. \emph{Base} represents the original model, while \emph{SVEN}~\cite{he2023large} and \emph{HexaCoder} refer to the models fine-tuned by each respective approach. For the CodeLMSec~\cite{hajipour2023systematically} and Pearce et al.~\cite{Pearce2022Asleep} benchmarks, the number of generated vulnerable codes is reported. For the HumanEval~\cite{Chen2021EvaluatingLL}, the pass@$10$ score is reported.
}
\vspace{-0.5em}
\label{hexacoder:table:overall:results}
\begin{adjustbox}{width=\textwidth,center}
\begin{tabular}{llll|lll|lll}
\toprule
Models    & \multicolumn{3}{c}{CodeGen-350M-multi~\cite{Nijkamp2022CG}}                                                                                                            & \multicolumn{3}{c}{Incoder-6B~\cite{fried2022incoder}}    &  \multicolumn{3}{c}{DeepSeek-Coder-V2-16B~\cite{zhu2024deepseekcoder2}}                             \\ \cmidrule(lr){1-1}\cmidrule(lr){2-4}\cmidrule(lr){5-7} \cmidrule(lr){8-10}
 & \multicolumn{1}{c}{CodeLMSec~\cite{hajipour2023systematically}}  & \multicolumn{1}{c}{Pearce et al.~\cite{Pearce2022Asleep}} & \multicolumn{1}{c}{HumanEval~\cite{Chen2021EvaluatingLL} } & \multicolumn{1}{c}{CodeLMSec~\cite{hajipour2023systematically}}  & \multicolumn{1}{c}{Pearce et al.~\cite{Pearce2022Asleep}} & \multicolumn{1}{c}{HumanEval~\cite{Chen2021EvaluatingLL}} & \multicolumn{1}{c}{CodeLMSec~\cite{hajipour2023systematically}}  & \multicolumn{1}{c}{Pearce et al.~\cite{Pearce2022Asleep}} & \multicolumn{1}{c}{HumanEval~\cite{Chen2021EvaluatingLL}}
           \\
 & \multicolumn{1}{c}{top-$5$ $\downarrow$}  & \multicolumn{1}{c}{top-$15$ $\downarrow$} & \multicolumn{1}{c}{pass@$10$ $\uparrow$} & \multicolumn{1}{c}{top-$5$ $\downarrow$}  & \multicolumn{1}{c}{top-$15$ $\downarrow$} & \multicolumn{1}{c}{pass@$10$ $\uparrow$} & \multicolumn{1}{c}{top-$5$ $\downarrow$}  & \multicolumn{1}{c}{top-$15$ $\downarrow$} & \multicolumn{1}{c}{pass@$10$ $\uparrow$ }
 \\
           \cmidrule(lr){2-4} \cmidrule(lr){5-7} \cmidrule(lr){8-10}
Base   &  \multicolumn{1}{c}{348} & \multicolumn{1}{c}{182} & \multicolumn{1}{c}{\textbf{9.9}} & \multicolumn{1}{c}{473} & \multicolumn{1}{c}{183} & \multicolumn{1}{c}{27.7} & \multicolumn{1}{c}{522} & \multicolumn{1}{c}{186}       & \multicolumn{1}{c}{70.5} \\

SVEN~\cite{he2023large}   &  \multicolumn{1}{c}{323} & \multicolumn{1}{c}{135} & \multicolumn{1}{c}{8.9} & \multicolumn{1}{c}{457} & \multicolumn{1}{c}{134} & \multicolumn{1}{c}{27.2} & \multicolumn{1}{c}{-} & \multicolumn{1}{c}{-}       & \multicolumn{1}{c}{-} \\

HexaCoder   &  \multicolumn{1}{c}{\textbf{105}} & \multicolumn{1}{c}{\textbf{55}} & \multicolumn{1}{c}{
8.4} & \multicolumn{1}{c}{\textbf{215}} & \multicolumn{1}{c}{\textbf{77}} & \multicolumn{1}{c}{\textbf{29.4}} & \multicolumn{1}{c}{\textbf{117}} & \multicolumn{1}{c}{\textbf{62}}       & \multicolumn{1}{c}{\textbf{72.0}} \\
\bottomrule
\end{tabular}
\end{adjustbox}
\vspace{-1.0em}
\end{table*}

\subsubsection{Effectiveness of the Two-step Generation Approach}

Our HexaCoder approach consists of fine-tuning the pre-trained models using the synthesized secure code data and generating the codes using the two-step generation approach. Here, we investigate the effectiveness of the two-step generation approach on the fine-tuned models with HexaCoder, as well as on models fine-tuned with SVEN~\cite{he2023large} and the original pre-trained model. Table~\ref{hexacoder:table:two-step:effect} shows the results of the number of vulnerable codes generated using different approaches, focusing on the top-$1$ and top-$5$ most probable samples. In this table, the CodeGen-350M-multi model serves as the base model in each case, with results provided both with and without the two-step generation approach. In Table~\ref{hexacoder:table:two-step:effect}, we compare the results for the pre-trained CodeGen-350M-multi~\cite{Nijkamp2022CG} (\textit{Base}), the fine-tuned version of the model using SVEN~\cite{he2023large}, and the fine-tuned model using our HexaCoder approach. In Table~\ref{hexacoder:table:two-step:effect}, \textit{Two} refers to our two-step approach. We provide the detailed results per each CWE in Appendix~\ref{hexacoder:appendix:subsection:results:detailed:two_step}.

In Table~\ref{hexacoder:table:two-step:effect}, we observe that our two-step generation for the pre-trained model (\textit{Base}) can even increase the number of vulnerable codes while for the fine-tuned model with SVEN~\cite{he2023large}, it reduces the number of vulnerable codes from 258 to 199 for top-5 most probable sample. This indicates that the number of vulnerable codes generated by SVEN with our two-step generation reduced by 22.8\% compared to the original SVEN approach. Table~\ref{hexacoder:table:two-step:effect} demonstrates that, among the top-5 samples, the fine-tuned model using HexaCoder generates 174 vulnerable codes without the two-step generation, whereas it only produces 81 vulnerable codes when our two-step generation approach is applied. This reflects a 53.4\% reduction in the number of vulnerable codes when using HexaCoder with the two-step generation compared to using HexaCoder without it. 

Table~\ref{hexacoder:table:two-step:effect} demonstrates that the two-step generation approach most effectively reduces the number of vulnerable codes in our method compared to the other approaches. This indicates that our synthesized dataset provides a better representation than SVEN's dataset. This is mainly due to the fact that our data synthesis pipeline includes the necessary libraries, while many samples in the SVEN dataset lack these essential components.
\begin{table}[tb]
\caption{Number of vulnerable Python code samples generated by the \textbf{CodeGen-350M-multi} model as evaluated using the prompt of the \textbf{CodeLMSec benchmark}, among the top-$1$ and top-$5$ most probable samples.  \emph{Base} represents the original model, while \emph{SVEN}~\cite{he2023large} and \emph{HexaCoder} refer to the CodeGen-350M-multi model fine-tuned by each respective approach. \textit{Two} denotes the two-step generation approach.}
\label{hexacoder:table:two-step:effect}

\centering
\begin{tabular}{cll}
\toprule
\multicolumn{1}{l}{Models} & top-$1$ & top-$5$\\
\cmidrule(lr){1-1} \cmidrule(lr){2-3}
\multicolumn{1}{l}{Base} &  59 & 281\\
\multicolumn{1}{l}{Base (w Two)} & 66 & 291\\
\cdashline{1-3}
\addlinespace
\multicolumn{1}{l}{SVEN} & 60 & 258\\
\multicolumn{1}{l}{SVEN (w Two)} & 45 & 199 \\
\cdashline{1-3}
\addlinespace
\multicolumn{1}{l}{HexaCoder (w/o Two)} & 41 & 174 \\
\multicolumn{1}{l}{HexaCoder} & 20 & 81\\
\bottomrule
\end{tabular}
\vspace{-0.4em}
\end{table}
\section{Discussion}
In this section, we discuss the limitations of our work, reflect on the lessons learned, and present potential future works.

\subsection{Limitations}

\paragraph{Static Analyzers.} In our work, we rely on CodeQL~\cite{codeql}, a static analysis tool, to identify the vulnerable and fixed codes. A static analyzers can only approximately identify software faults and does not guarantee sound results~\cite{chess2004static,10.1145/2644805}. Following previous work~\cite{Pearce2022Asleep, he2023large,hajipour2023systematically,he2024safecoder}, we decided to use CodeQL~\cite{codeql}, which is one of the best-performing and freely available tools covering a wide range of CWEs~\cite{hajipour2023systematically,lipp2022empirical}. 

\paragraph{Programming Languages.} We demonstrate the effectiveness of our approach using the programming languages Python and C/C++. These languages were chosen because they are widely used in the community, and the CodeLMSec~\cite{hajipour2023systematically} and Pearce et al.~\cite{Pearce2022Asleep} benchmarks provide prompts in these languages for evaluating code security aspects of models. In future work, it would be valuable to expand the set of programming languages to include other languages, such as JavaScript, PHP, and Go.

\paragraph{Funtionality Changes in the Fixed Codes.} In our data synthesis pipeline, we use the GPT-4 model~\cite{openai2023gpt4} to generate the fixed codes based on the provided security report and the vulnerable code. One concern is that the output might produce code with different functionality or even empty code. However, our observations show that the model's output typically includes one or more changes aimed at resolving the security issue. Although there is no guarantee that the output code will maintain the exact same functionality, our results demonstrate that by using our synthesized data, the models are able to generate code with fewer vulnerabilities while preserving utility performance. Note that in our data synthesis pipeline, we do not consider the empty output as fixed codes.

\paragraph{Potential Issues of the Two-Step Generation Approach.} In our experimental results, we show the effectiveness of our two-step generation approach in reducing the number of vulnerable codes. However, in this approach, the first step of generating new libraries based on the given prompt does not take the context of the code into account. As a result, unused libraries may be generated. This issue can be resolved by removing these unused libraries from the code. Furthermore, the performance of our approach in generating functionally correct codes shows that with the two-step generation approach, we can achieve reasonable performance. Note that we also tried to incorporate the context of the prompt in the first step of the code generation using fill-in-the-middle generation fashion~\cite{zhu2024deepseekcoder2}. However, we found that this method did not yield improvements as significant as those achieved by generating the library in an autoregressive fashion.

\subsection{Adaptability of Our Approach to Other Vulnerabilities}
In our data synthesis pipeline, we use the few-shot prompting method introduced in~\cite{hajipour2023systematically} to generate vulnerable code samples. This technique requires only a few examples of code with the targeted vulnerability to create additional instances of similar vulnerable code. We then leverage a security oracle alongside GPT-4~\cite{openai2023gpt4} to correct these vulnerabilities. Consequently, our method can be easily adapted to address other types of vulnerabilities as well. Exploring the application of our data synthesis pipeline to other and new types of vulnerabilities presents an interesting research direction.

\section{Conclusion}
In this paper, we presented HexaCoder, a novel approach to enhancing the security of code generated by CodeLMs. HexaCoder 
consists of an oracle-guided data synthesis pipeline and a two-step code generation process. Using the end-to-end data synthesis pipeline, HexaCoder generates pairs of vulnerable and fixed codes for targeted CWEs and uses these codes to fine-tune the CodeLMs using the LoRA method. During inference, the proposed two-step generation approach allows the trained models to include all necessary libraries that may have been initially overlooked, thus enabling the generation of more secure code. As a result, this approach significantly reduces the occurrence of vulnerable code in the output. Our comprehensive evaluation across three different benchmarks and four CodeLMs demonstrates that HexaCoder not only improves the security of generated code but also preserves its functional correctness, addressing a critical balance in the field of automatic code generation.

\clearpage

\section*{Acknowledgments}
This work was partially funded by ELSA – European Lighthouse on Secure
and Safe AI funded by the European Union under grant agreement No.
101070617. Views and opinions expressed are however those of the authors
only and do not necessarily reflect those of the European Union or European
Commission. Neither the European Union nor the European Commission
can be held responsible for them.
This work was partially funded by the German Federal Ministry of Education and Research (BMBF) under the grant AIgenCY (16KIS2012).

%\clearpage
% \printbibliography
\Urlmuskip=0mu plus 1mu\relax
\bibliographystyle{plain}
\bibliography{main}

\begin{thebibliography}{10}

\bibitem{tabnine}
{Tabnine}, 2013.

\bibitem{codeium}
{Codeium}, 2023.

\bibitem{ghostwriter}
{Ghostwriter - Code faster with AI}, 2023.

\bibitem{codeiumai}
{Codium AI}, 2024.

\bibitem{abay2019privacy}
Nazmiye~Ceren Abay, Yan Zhou, Murat Kantarcioglu, Bhavani Thuraisingham, and Latanya Sweeney.
\newblock Privacy preserving synthetic data release using deep learning.
\newblock In {\em ECML PKDD}, 2019.

\bibitem{asare2023github}
Owura Asare, Meiyappan Nagappan, and N~Asokan.
\newblock Is github’s copilot as bad as humans at introducing vulnerabilities in code?
\newblock {\em Empirical Software Engineering}, 2023.

\bibitem{4602670}
Nathaniel Ayewah, William Pugh, David Hovemeyer, J.~David Morgenthaler, and John Penix.
\newblock Using static analysis to find bugs.
\newblock {\em IEEE Software}, 2008.

\bibitem{babbar2019data}
Rohit Babbar and Bernhard Sch{\"o}lkopf.
\newblock Data scarcity, robustness and extreme multi-label classification.
\newblock {\em Machine Learning}, 2019.

\bibitem{7476667}
Moritz Beller, Radjino Bholanath, Shane McIntosh, and Andy Zaidman.
\newblock Analyzing the state of static analysis: A large-scale evaluation in open source software.
\newblock In {\em SANER}, 2016.

\bibitem{bhatt2023purple}
Manish Bhatt, Sahana Chennabasappa, Cyrus Nikolaidis, Shengye Wan, Ivan Evtimov, Dominik Gabi, Daniel Song, Faizan Ahmad, Cornelius Aschermann, Lorenzo Fontana, et~al.
\newblock Purple llama cyberseceval: A secure coding benchmark for language models.
\newblock {\em arXiv}, 2023.

\bibitem{biderman2024lora}
Dan Biderman, Jose~Gonzalez Ortiz, Jacob Portes, Mansheej Paul, Philip Greengard, Connor Jennings, Daniel King, Sam Havens, Vitaliy Chiley, Jonathan Frankle, et~al.
\newblock Lora learns less and forgets less.
\newblock {\em arXiv}, 2024.

\bibitem{brown2020language}
Tom Brown, Benjamin Mann, Nick Ryder, Melanie Subbiah, Jared~D Kaplan, Prafulla Dhariwal, Arvind Neelakantan, Pranav Shyam, Girish Sastry, Amanda Askell, et~al.
\newblock Language models are few-shot learners.
\newblock In {\em NeurIPS}, 2020.

\bibitem{cassano2022multipl}
Federico Cassano, John Gouwar, Daniel Nguyen, Sydney Nguyen, Luna Phipps-Costin, Donald Pinckney, Ming-Ho Yee, Yangtian Zi, Carolyn~Jane Anderson, Molly~Q Feldman, et~al.
\newblock Multipl-e: A scalable and extensible approach to benchmarking neural code generation.
\newblock {\em arXiv}, 2022.

\bibitem{chang2024survey}
Hsin-Yu Chang, Pei-Yu Chen, Tun-Hsiang Chou, Chang-Sheng Kao, Hsuan-Yun Yu, Yen-Ting Lin, and Yun-Nung Chen.
\newblock A survey of data synthesis approaches.
\newblock {\em arXiv}, 2024.

\bibitem{codealpaca}
Sahil Chaudhary.
\newblock Code alpaca: An instruction-following llama model for code generation.
\newblock \url{https://github.com/sahil280114/codealpaca}, 2023.

\bibitem{Chen2021EvaluatingLL}
Mark Chen, Jerry Tworek, Heewoo Jun, Qiming Yuan, Henrique Ponde, Jared Kaplan, Harrison Edwards, Yura Burda, Nicholas Joseph, Greg Brockman, Alex Ray, Raul Puri, Gretchen Krueger, Michael Petrov, Heidy Khlaaf, Girish Sastry, Pamela Mishkin, Brooke Chan, Scott Gray, Nick Ryder, Mikhail Pavlov, Alethea Power, Lukasz Kaiser, Mohammad Bavarian, Clemens Winter, Philippe Tillet, Felipe~Petroski Such, David~W. Cummings, Matthias Plappert, Fotios Chantzis, Elizabeth Barnes, Ariel Herbert-Voss, William~H. Guss, Alex Nichol, Igor Babuschkin, S.~Arun Balaji, Shantanu Jain, Andrew Carr, Jan Leike, Joshua Achiam, Vedant Misra, Evan Morikawa, Alec Radford, Matthew~M. Knight, Miles Brundage, Mira Murati, Katie Mayer, Peter Welinder, Bob McGrew, Dario Amodei, Sam McCandlish, Ilya Sutskever, and Wojciech Zaremba.
\newblock Evaluating large language models trained on code.
\newblock {\em arXiv}, 2021.

\bibitem{chess2004static}
Brian Chess and Gary McGraw.
\newblock Static analysis for security.
\newblock {\em IEEE Security \& Privacy}, 2(6), 2004.

\bibitem{10.1145/2970276.2970347}
Maria Christakis and Christian Bird.
\newblock What developers want and need from program analysis: An empirical study.
\newblock In {\em IEEE/ACM International Conference on Automated Software Engineering (ASE)}, 2016.

\bibitem{deepseekv2}
DeepSeek-AI.
\newblock Deepseek-v2: A strong, economical, and efficient mixture-of-experts language model.
\newblock {\em arXiv}, 2024.

\bibitem{devlin-etal-2019-bert}
Jacob Devlin, Ming-Wei Chang, Kenton Lee, and Kristina Toutanova.
\newblock {BERT}: Pre-training of deep bidirectional transformers for language understanding.
\newblock In {\em NAACL}, 2019.

\bibitem{difflib2023}
{difflib}.
\newblock difflib - helpers for computing deltas, 2023.
\newblock {\url{https://docs.python.org/3/library/difflib.html}, as of \today}.

\bibitem{github-22-copilot}
Thomas Dohmke.
\newblock Github copilot is generally available to all developers, June 2022.
\newblock \url{https://github.blog/2022-06-21-github-copilot-is-generally-available-to-all-developers/}, as of \today.

\bibitem{dubey2024llama}
Abhimanyu Dubey, Abhinav Jauhri, Abhinav Pandey, Abhishek Kadian, Ahmad Al-Dahle, Aiesha Letman, Akhil Mathur, Alan Schelten, Amy Yang, Angela Fan, et~al.
\newblock The llama 3 herd of models.
\newblock {\em arXiv}, 2024.

\bibitem{codebert}
Zhangyin Feng, Daya Guo, Duyu Tang, Nan Duan, Xiaocheng Feng, Ming Gong, Linjun Shou, Bing Qin, Ting Liu, Daxin Jiang, and Ming Zhou.
\newblock {C}ode{BERT}: A pre-trained model for programming and natural languages.
\newblock In {\em EMNLP}, 2020.

\bibitem{fioraldi-20-afl++}
Andrea Fioraldi, Dominik Maier, Heiko Eißfeldt, and Marc Heuse.
\newblock {AFL++ : Combining Incremental Steps of Fuzzing Research }.
\newblock In {\em WOOT}, 2020.

\bibitem{fioraldi2022libafl}
Andrea Fioraldi, Dominik~Christian Maier, Dongjia Zhang, and Davide Balzarotti.
\newblock Libafl: A framework to build modular and reusable fuzzers.
\newblock In {\em CCS}, 2022.

\bibitem{fried2022incoder}
Daniel Fried, Armen Aghajanyan, Jessy Lin, Sida Wang, Eric Wallace, Freda Shi, Ruiqi Zhong, Wen-tau Yih, Luke Zettlemoyer, and Mike Lewis.
\newblock Incoder: A generative model for code infilling and synthesis.
\newblock {\em arXiv}, 2022.

\bibitem{gao2023scaling}
Leo Gao, John Schulman, and Jacob Hilton.
\newblock Scaling laws for reward model overoptimization.
\newblock In {\em ICML}, 2023.

\bibitem{gilardi2023chatgpt}
Fabrizio Gilardi, Meysam Alizadeh, and Ma{\"e}l Kubli.
\newblock Chatgpt outperforms crowd workers for text-annotation tasks.
\newblock {\em National Academy of Sciences}, 2023.

\bibitem{graphcodebert}
Daya Guo, Shuo Ren, Shuai Lu, Zhangyin Feng, Duyu Tang, Shujie Liu, Long Zhou, Nan Duan, Alexey Svyatkovskiy, Shengyu Fu, Michele Tufano, Shao~Kun Deng, Colin~B. Clement, Dawn Drain, Neel Sundaresan, Jian Yin, Daxin Jiang, and Ming Zhou.
\newblock Graphcodebert: Pre-training code representations with data flow.
\newblock In {\em ICLR}, 2021.

\bibitem{guo2024deepseekcoder}
Daya Guo, Qihao Zhu, Dejian Yang, Zhenda Xie, Kai Dong, Wentao Zhang, Guanting Chen, Xiao Bi, Yu~Wu, YK~Li, et~al.
\newblock Deepseek-coder: When the large language model meets programming--the rise of code intelligence.
\newblock {\em arXiv}, 2024.

\bibitem{hajipour2023systematically}
Hossein Hajipour, Hassler Keno, Thorsten Holz, Lea Sch{\"o}nherr, and Mario Fritz.
\newblock Codelmsec benchmark: Systematically evaluating and finding security vulnerabilities in black-box code language models.
\newblock In {\em IEEE SaTML}, 2024.

\bibitem{hajipour24simscood}
Hossein Hajipour, Ning Yu, Cristian-Alexandru Staicu, and Mario Fritz.
\newblock Simscood: Systematic analysis of out-of-distribution generalization in fine-tuned source code models.
\newblock In {\em NAACL Findings}, 2024.

\bibitem{hamer2024just}
Sivana Hamer, Marcelo d’Amorim, and Laurie Williams.
\newblock Just another copy and paste? comparing the security vulnerabilities of chatgpt generated code and stackoverflow answers.
\newblock In {\em IEEE Symposium on Security and Privacy Workshops}, 2024.

\bibitem{he2023large}
Jingxuan He and Martin Vechev.
\newblock Large language models for code: Security hardening and adversarial testing.
\newblock In {\em CCS}, 2023.

\bibitem{he2024safecoder}
Jingxuan He, Mark Vero, Gabriela Krasnopolska, and Martin Vechev.
\newblock Instruction tuning for secure code generation.
\newblock In {\em ICML}, 2024.

\bibitem{pmlr-v97-houlsby19a}
Neil Houlsby, Andrei Giurgiu, Stanislaw Jastrzebski, Bruna Morrone, Quentin De~Laroussilhe, Andrea Gesmundo, Mona Attariyan, and Sylvain Gelly.
\newblock Parameter-efficient transfer learning for {NLP}.
\newblock In {\em ICML}, 2019.

\bibitem{hu2022lora}
Edward~J Hu, yelong shen, Phillip Wallis, Zeyuan Allen-Zhu, Yuanzhi Li, Shean Wang, Lu~Wang, and Weizhu Chen.
\newblock Lo{RA}: Low-rank adaptation of large language models.
\newblock In {\em ICLR}, 2022.

\bibitem{codeql}
GitHub Inc.
\newblock Github codeql, 2022.
\newblock \url{https://codeql.github.com/}, as of \today.

\bibitem{semgrep}
Semgrep Inc.
\newblock {Find bugs and reachable dependency vulnerabilities in code.}, 2024.

\bibitem{khoury2023secure}
Rapha{\"e}l Khoury, Anderson~R Avila, Jacob Brunelle, and Baba~Mamadou Camara.
\newblock How secure is code generated by chatgpt?
\newblock In {\em IEEE SMC}, 2023.

\bibitem{li2023starcoder}
Raymond Li, Loubna~Ben Allal, Yangtian Zi, Niklas Muennighoff, Denis Kocetkov, Chenghao Mou, Marc Marone, Christopher Akiki, Jia Li, Jenny Chim, et~al.
\newblock Starcoder: may the source be with you!
\newblock {\em TMLR}, 2023.

\bibitem{lipp2022empirical}
Stephan Lipp, Sebastian Banescu, and Alexander Pretschner.
\newblock An empirical study on the effectiveness of static c code analyzers for vulnerability detection.
\newblock In {\em ISSTA}, pages 544--555, 2022.

\bibitem{liu2024best}
Ruibo Liu, Jerry Wei, Fangyu Liu, Chenglei Si, Yanzhe Zhang, Jinmeng Rao, Steven Zheng, Daiyi Peng, Diyi Yang, Denny Zhou, et~al.
\newblock Best practices and lessons learned on synthetic data for language models.
\newblock {\em arXiv}, 2024.

\bibitem{10.1145/2644805}
Benjamin Livshits, Manu Sridharan, Yannis Smaragdakis, Ond\v{r}ej Lhot\'{a}k, J.~Nelson Amaral, Bor-Yuh~Evan Chang, Samuel~Z. Guyer, Uday~P. Khedker, Anders M\o{}ller, and Dimitrios Vardoulakis.
\newblock In defense of soundiness: a manifesto.
\newblock {\em Communication of the ACM}, 58(2), 2015.

\bibitem{long2024llms}
Lin Long, Rui Wang, Ruixuan Xiao, Junbo Zhao, Xiao Ding, Gang Chen, and Haobo Wang.
\newblock On llms-driven synthetic data generation, curation, and evaluation: A survey.
\newblock {\em arXiv}, 2024.

\bibitem{lozhkov2024starcoder}
Anton Lozhkov, Raymond Li, Loubna~Ben Allal, Federico Cassano, Joel Lamy-Poirier, Nouamane Tazi, Ao~Tang, Dmytro Pykhtar, Jiawei Liu, Yuxiang Wei, et~al.
\newblock Starcoder 2 and the stack v2: The next generation.
\newblock {\em arXiv}, 2024.

\bibitem{luo2023wizardmath}
Haipeng Luo, Qingfeng Sun, Can Xu, Pu~Zhao, Jianguang Lou, Chongyang Tao, Xiubo Geng, Qingwei Lin, Shifeng Chen, and Dongmei Zhang.
\newblock Wizardmath: Empowering mathematical reasoning for large language models via reinforced evol-instruct.
\newblock {\em arXiv}, 2023.

\bibitem{luo2024wizardcoder}
Ziyang Luo, Can Xu, Pu~Zhao, Qingfeng Sun, Xiubo Geng, Wenxiang Hu, Chongyang Tao, Jing Ma, Qingwei Lin, and Daxin Jiang.
\newblock Wizardcoder: Empowering code large language models with evol-instruct.
\newblock In {\em ICLR}, 2024.

\bibitem{meng2023tuning}
Yu~Meng, Martin Michalski, Jiaxin Huang, Yu~Zhang, Tarek Abdelzaher, and Jiawei Han.
\newblock Tuning language models as training data generators for augmentation-enhanced few-shot learning.
\newblock In {\em ICML}, 2023.

\bibitem{mitre}
MITRE.
\newblock {CWE} - {C}ommon {W}eakness {E}numeration, 2022.
\newblock \url{https://cwe.mitre.org}, as of \today.

\bibitem{Nijkamp2022CG}
Erik Nijkamp, Bo~Pang, Hiroaki Hayashi, Lifu Tu, Huan Wang, Yingbo Zhou, Silvio Savarese, and Caiming Xiong.
\newblock Codegen: An open large language model for code with multi-turn program synthesis.
\newblock In {\em ICLR}, 2023.

\bibitem{openai-22-chatgpt}
OpenAI.
\newblock Chatgpt: Optimizing language models for dialogue, November 2022.
\newblock \url{https://openai.com/blog/chatgpt/}, as of \today.

\bibitem{openai2023gpt4}
OpenAI.
\newblock Gpt-4 technical report, 2023.

\bibitem{or2019dynamic}
Ori Or-Meir, Nir Nissim, Yuval Elovici, and Lior Rokach.
\newblock Dynamic malware analysis in the modern era: A state of the art survey.
\newblock {\em ACM Computing Surveys (CSUR)}, 2019.

\bibitem{Pearce2022Asleep}
Hammond Pearce, Baleegh Ahmad, Benjamin Tan, Brendan Dolan-Gavitt, and Ramesh Karri.
\newblock Asleep at the keyboard? assessing the security of github copilot's code contributions.
\newblock In {\em IEEE Symposium on Security and Privacy}, 2022.

\bibitem{radford2019language}
Alec Radford, Jeffrey Wu, Rewon Child, David Luan, Dario Amodei, Ilya Sutskever, et~al.
\newblock Language models are unsupervised multitask learners.
\newblock {\em OpenAI blog}, 2019.

\bibitem{codellama}
Baptiste Rozi{\`e}re, Jonas Gehring, Fabian Gloeckle, Sten Sootla, Itai Gat, Xiaoqing~Ellen Tan, Yossi Adi, Jingyu Liu, Tal Remez, J{\'e}r{\'e}my Rapin, et~al.
\newblock Code llama: Open foundation models for code.
\newblock {\em arXiv}, 2023.

\bibitem{sandoval2022security}
Gustavo Sandoval, Hammond Pearce, Teo Nys, Ramesh Karri, Brendan Dolan-Gavitt, and Siddharth Garg.
\newblock Security implications of large language model code assistants: A user study.
\newblock {\em arXiv}, 2022.

\bibitem{7546500}
Yan Shoshitaishvili, Ruoyu Wang, Christopher Salls, Nick Stephens, Mario Polino, Andrew Dutcher, John Grosen, Siji Feng, Christophe Hauser, Christopher Kruegel, and Giovanni Vigna.
\newblock {SoK: (State of) The Art of War: Offensive Techniques in Binary Analysis}.
\newblock In {\em IEEE Symposium on Security and Privacy}, 2016.

\bibitem{siddiq2022securityeval}
Mohammed~Latif Siddiq and Joanna~CS Santos.
\newblock Securityeval dataset: mining vulnerability examples to evaluate machine learning-based code generation techniques.
\newblock In {\em MSR4 P and S}, 2022.

\bibitem{6547101}
László Szekeres, Mathias Payer, Tao Wei, and Dawn Song.
\newblock {SoK: Eternal War in Memory}.
\newblock In {\em IEEE Symposium on Security and Privacy}, 2013.

\bibitem{tihanyi2023formai}
Norbert Tihanyi, Tamas Bisztray, Ridhi Jain, Mohamed~Amine Ferrag, Lucas~C Cordeiro, and Vasileios Mavroeidis.
\newblock The formai dataset: Generative ai in software security through the lens of formal verification.
\newblock In {\em PROMISE}, 2023.

\bibitem{wang-etal-2023-self-instruct}
Yizhong Wang, Yeganeh Kordi, Swaroop Mishra, Alisa Liu, Noah~A. Smith, Daniel Khashabi, and Hannaneh Hajishirzi.
\newblock Self-instruct: Aligning language models with self-generated instructions.
\newblock In {\em ACL}, 2023.

\bibitem{codet5}
Yue Wang, Weishi Wang, Shafiq Joty, and Steven~C.H. Hoi.
\newblock Codet5: Identifier-aware unified pre-trained encoder-decoder models for code understanding and generation.
\newblock In {\em EMNLP}, 2021.

\bibitem{wei2023magicoder}
Yuxiang Wei, Zhe Wang, Jiawei Liu, Yifeng Ding, and Lingming Zhang.
\newblock Magicoder: Source code is all you need.
\newblock {\em arXiv}, 2023.

\bibitem{wiki:Markdown}
Wikipedia.
\newblock {Markdown} --- {W}ikipedia{,} the free encyclopedia, 2024.
\newblock \url{http://en.wikipedia.org/w/index.php?title=Markdown&oldid=1241689312}, as of \today.

\bibitem{xu2024wizardlm}
Can Xu, Qingfeng Sun, Kai Zheng, Xiubo Geng, Pu~Zhao, Jiazhan Feng, Chongyang Tao, Qingwei Lin, and Daxin Jiang.
\newblock Wizard{LM}: Empowering large pre-trained language models to follow complex instructions.
\newblock In {\em ICLR}, 2024.

\bibitem{yu2023metamath}
Longhui Yu, Weisen Jiang, Han Shi, Jincheng Yu, Zhengying Liu, Yu~Zhang, James~T Kwok, Zhenguo Li, Adrian Weller, and Weiyang Liu.
\newblock Metamath: Bootstrap your own mathematical questions for large language models.
\newblock {\em arXiv}, 2023.

\bibitem{github-22-copilot-biz}
Shuyin Zhao.
\newblock Github copilot is generally available for businesses, December 2022.
\newblock \url{https://github.blog/2022-12-07-github-copilot-is-generally-available-for-businesses/}, as of \today.

\bibitem{zhu2024deepseekcoder2}
Qihao Zhu, Daya Guo, Zhihong Shao, Dejian Yang, Peiyi Wang, Runxin Xu, Y~Wu, Yukun Li, Huazuo Gao, Shirong Ma, et~al.
\newblock Deepseek-coder-v2: Breaking the barrier of closed-source models in code intelligence.
\newblock {\em arXiv}, 2024.

\bibitem{zhu2022fuzzing}
Xiaogang Zhu, Sheng Wen, Seyit Camtepe, and Yang Xiang.
\newblock Fuzzing: a survey for roadmap.
\newblock {\em ACM Computing Surveys (CSUR)}, 54(11s), 2022.

\end{thebibliography}
%%%%%%%%%%%%%%%%%%%%%%%%%%%%%%%%%%%%%%%%%%%%%%%%%%%%%%%%%%%%%%%%%%%%%%%%%%%%%%%%

\clearpage

\appendix
\section{Details of the Input prompts}
\label{hexacoder:appendix:prompts}

Figure~\ref{hexacoder:appendix:fig:full_template} provides the full version of the input prompt's template that we demonstrated in Figure~\ref{hexacoder:fig:template}. Here, we provide the full list of instructions that we used in our synthesis pipeline. 

In Section~\ref{hexacoder:subsection:performance:code_syn}, we compared three variations of the secure code synthesis approach. In one of these variations, we do not use any security report. In Figure~\ref{hexacoder:appendix:fig:template_without_sec_report}, we provide the input prompt that we used for this specific case.

\begin{figure}[h]
  \vspace{-1.5em}
  \centering
\descriptionbox{
\texttt{You are a security engineer and \textbf{\{prog\_lang\}} developer. The following code has \textbf{\{num\_vuls\}} vulnerability(ies):}
% \vspace{1em}

\texttt{\textbf{\{vul\_count\}}- The code has a CWE vulnerability at line \textbf{\texttt{\{line\_num\}}}.
The vulnerability is of \textbf{\texttt{\{cwe\_type\}}} type (\textbf{\texttt{\{cwe\_explanation\}}}).}
\vspace{1em}

\textbf{\texttt{\{hint\}}}

\vspace{1em}
\texttt{\textbf{Instructions:}
\begin{enumerate}
    \item \textbf{Analysis}: First, provide a detailed explanation of the vulnerabilities present. Describe the steps necessary to fix these issues.
    \item \textbf{Correction}: After your explanation, directly repair the code. Ensure the following:
    \begin{itemize}
        \item Correct all vulnerabilities in a single solution.
        \item Avoid any syntax errors and ensure the code is valid in \texttt{\textbf{\{prog\_lang\}}}.
        \item Do not provide multiple solutions or additional commentary after the corrected code.
        \item Present the repaired code in a Markdown code block for \texttt{\textbf{\{prog\_lang\}}}.
        \item Do not write any explanation after the corrected code.
        \item If new libraries are required, include them after the current included libraries.
    \end{itemize}
\end{enumerate}
}
\texttt{Vulnerable code:} \\
``` \texttt{\textbf{\{prog\_lang\}}} \\
\texttt{\textbf{\{vul\_code\}}} '''\\

\texttt{\textbf{Expected Outputs:}}
\begin{itemize}
    \item \texttt{A clear and concise description of how to address the vulnerabilities. This is a MUST.}
    \item \texttt{The corrected version of the code.}
\end{itemize}
}

\vspace{-1.0em}
\caption{Template of the input prompt.}
  \vspace{-1.0em}
  \label{hexacoder:appendix:fig:full_template}
\end{figure}

\begin{figure}[h]
  \vspace{-1.0em}
  \centering
\descriptionbox{
\texttt{You are a \textbf{\{prog\_lang\}} developer.}

\vspace{1em}
\texttt{\textbf{Instructions:}
\begin{enumerate}
    \item \textbf{Analysis}:If there is a vulnerability in the code, provide a detailed explanation of the vulnerabilities present. Describe the steps necessary to fix these issues.
    \item \textbf{Correction}: If there is a vulnerability in the code, repair the code. Ensure the following:
    \begin{itemize}
        \item Correct all vulnerabilities in a single solution.
        \item Avoid any syntax errors and ensure the code is valid in \texttt{\textbf{\{prog\_lang\}}}.
        \item Do not provide multiple solutions or additional commentary after the corrected code.
        \item Present the repaired code in a Markdown code block for \texttt{\textbf{\{prog\_lang\}}}.
        \item Do not write any explanation after the corrected code.
        \item If new libraries are required, include them after the current included libraries.
    \end{itemize}
\end{enumerate}
}
\texttt{Code:} \\
``` \texttt{\textbf{\{prog\_lang\}}} \\
\texttt{\textbf{\{vul\_code\}}} '''\\

\texttt{\textbf{Expected Outputs:}}
\begin{itemize}
    \item \texttt{A clear and concise description of how to address the vulnerabilities. This is a MUST.}
    \item \texttt{The corrected version of the code.}
\end{itemize}
}

\vspace{-1.0em}
\caption{Template of the input prompt without using the security report.}
  \vspace{-1.0em}
  \label{hexacoder:appendix:fig:template_without_sec_report}
\end{figure}

\section{Details of the Security Hints}
\label{hexacoder:appendix:hints}
In our code synthesis pipeline, we provide security hints to guide the model in resolving the security vulnerabilities in the given code. In Table~\ref{hexacoder:appendix:table:hints}, we provide the list of the security hints that we employed in our work. 

\begin{table*}[tb]
\caption{List of the security hints that have been used in the security report of our code synthesis pipeline.}
\label{hexacoder:appendix:table:hints}

\centering
\begin{tabular}{lp{0.8\textwidth}}
\toprule
CWE & Hint \\
\cmidrule(lr){1-1} \cmidrule(lr){2-2} 
CWE-020 & Data from all potentially untrusted sources should be subject to input validation. \\
\cdashline{1-2}
\addlinespace

CWE-022-Py & Inputs should be decoded and canonicalized to the application's current internal representation before being validated. Sanitize the user's input using safe\_join or os.path.normpath. \\
\cdashline{1-2}
\addlinespace

CWE-022-C/C++ & Inputs should be decoded and canonicalized to the application's current internal representation before being validated. Use a built-in path canonicalization function (such as realpath() in C) that produces the canonical version of the pathname, which effectively removes sequences and symbolic links (CWE-23, CWE-59). \\
\cdashline{1-2}
\addlinespace

CWE-078 & Properly quote arguments and escape any special characters within those arguments. If using subprocess, avoid using shell=True. Alternatively, use shell=False instead. \\
\cdashline{1-2}
\addlinespace

CWE-079 & Note that proper output encoding, escaping, and quoting is the most effective solution for preventing XSS. \\
\cdashline{1-2}
\addlinespace

CWE-094 & Avoid using eval, exec, execfile functions or validate the user input. Use the ast.literal\_eval() function. However, consider that an adversary could still cause excessive memory or stack consumption via deeply nested structures. \\
\cdashline{1-2}
\addlinespace
CWE-117 &  User input should be suitably sanitized before it is logged. If the log entries are plain text, then line breaks should be removed from user input, such as replacing (old, new) or similar. Care should also be taken that user input is clearly marked in log entries, and that a malicious user cannot cause confusion in other ways.For log entries that will be displayed in HTML, user input should be HTML encoded before being logged, to prevent forgery and other forms of HTML injection. \\
\cdashline{1-2}
\addlinespace
CWE-190 & Perform input validation on any numeric input by ensuring that it is within the expected range. Enforce that the input meets both the minimum and maximum requirements for the expected range. Use unsigned integers where possible. This makes it easier to perform validation for integer overflows. When signed integers are required, ensure that the range check includes minimum values as well as maximum values. \\
\cdashline{1-2}
\addlinespace
CWE-476 & If all pointers that could have been modified are sanity-checked previous to use, nearly all NULL pointer dereferences can be prevented. \\
\cdashline{1-2}
\addlinespace
CWE-502 & When deserializing data, ensure that a new object is populated rather than just deserialized into an existing one. This ensures that the data flows through safe input validation processes and that the functions remain secure. \\
\cdashline{1-2}
\addlinespace
CWE-611 & Many XML parsers and validators can be configured to disable external entity expansion. It is recommended to use the defusedxml library instead of the native Python XML libraries. The defusedxml library is specifically designed to mitigate XML external entity attacks. To guard against XXE attacks with the lxml library, you should create a parser with resolve\_entities set to false. \\
\cdashline{1-2}
\addlinespace
CWE-787 & Consider adhering to the following rules when allocating and managing an application's memory:\begin{itemize}
    \item Double check that the buffer is as large as specified.
    \item When using functions that accept a number of bytes to copy, such as strncpy(), be aware that if the destination buffer size is equal to the source buffer size, it may not NULL-terminate the string.
    \item Check buffer boundaries if accessing the buffer in a loop and make sure there is no danger of writing past the allocated space.
    \item If necessary, truncate all input strings to a reasonable length before passing them to the copy and concatenation functions.
\end{itemize} \\
\bottomrule
\end{tabular}
% \vspace{-0.2cm}
\end{table*}

\section{Hyperparameters for LoRA Fine-tuning Approach}
\label{hexacoder:appendix:lora}
In Table~\ref{hexacoder:appendix:table:lora}, we list the LoRA hyperparameters used to fine-tune the various CodeLMs. We aimed to keep most parameters consistent across models. However, due to computational constraints, we set the LoRA rank for DeepSeek-Coder-V2-16B~\cite{zhu2024deepseekcoder2} to 16, while for the other models, the rank was set to 64.
\begin{table*}[h]

\begin{center}
\caption{The LoRA hyperparameters we used to fine-tune the pre-trained CodeLMs.}
\label{hexacoder:appendix:table:lora}
\end{center}
\begin{center}
\begin{tabular}{lccccc}
\toprule
\multicolumn{1}{c}{Models} & \multicolumn{1}{l}{Batch Size} & \#Epoch & Learning Rate & Rank & LoRA $\alpha$ \\ \hline
CodeGen-350M-multi~\cite{Nijkamp2022CG}                     & 16                             & 10      & $5e^{-4}$               & 64   & 16     \\
CodeGen-2B-multi~\cite{Nijkamp2022CG}                     & 16                             & 10      &  $5e^{-4}$              & 64   & 16     \\
Incoder-6B~\cite{fried2022incoder}                    & 16                             & 10      &  $5e^{-4}$              & 64   & 16     \\
DeepSeek-Coder-V2-16B~\cite{zhu2024deepseekcoder2}                 & 16                              & 10       &  $5e^{-4}$              & 16   & 16    
\\
\bottomrule

\end{tabular}

\end{center}
\end{table*}

\section{Detailed Results}

Tables~\ref{hexacoder:appendix:table:codelmsec:codegen350},~\ref{hexacoder:appendix:table:pearce:codegen350m},\ref{hexacoder:appendix:table:codelmsec:incoder},~\ref{hexacoder:appendix:table:pearce:incoder},~\ref{hexacoder:appendix:table:codelmsec:deepseek}, and~\ref{hexacoder:appendix:table:pearce:deepseek} provide the detailed results of evaluating different variations of the models using CodeLMSec~\cite{hajipour2023systematically} and Pearce et al.~\cite{Pearce2022Asleep} benchmarks. These tables provide the number of generated vulnerable codes per CWE for Python and C/C++ codes.

Tables~\ref{hexacoder:appendix:table:codelmsec:codegen350} and~\ref{hexacoder:appendix:table:pearce:codegen350m} demonstrate the number of vulnerable codes generated by different variations of CodeGen-350M-multi~\cite{Nijkamp2022CG}. In Table ~\ref{hexacoder:appendix:table:codelmsec:codegen350}, we observe that the CodeGen-350M-multi model fine-tuned with our HexaCoder produces fewer vulnerable Python codes compared to the other variations of the model. For example, using our approach, the model generates no code containing CWE-078 vulnerabilities, whereas the pre-trained model generates 12 such instances, and SVEN~\cite{he2023large} generates 10. Table~\ref{hexacoder:appendix:table:pearce:codegen350m} also shows that in 9 out of 13 cases, our approach generates the same or fewer number of vulnerable codes than the other approaches.

In Tables~\ref{hexacoder:appendix:table:codelmsec:incoder} and~\ref{hexacoder:appendix:table:pearce:incoder} we provide the number of vulnerable codes generated by different variations of InCoder-6B~\cite{fried2022incoder}. These tables also demonstrate that fine-tuning InCoder-6B~\cite{fried2022incoder} using our approach significantly reduces the number of vulnerable codes that can be generated using CodeLMSec~\cite{hajipour2023systematically} and Pearce et al.~\cite{Pearce2022Asleep} benchmarks. 

Tables~\ref{hexacoder:appendix:table:codelmsec:deepseek} and~\ref{hexacoder:appendix:table:pearce:deepseek} demonstrate the number of vulnerable codes generated by different variations of DeepSeek-Coder-V2-16B~\cite{zhu2024deepseekcoder2}. In these two tables, as the fine-tuned version of DeepSeek-Coder-V2-16B~\cite{zhu2024deepseekcoder2} using SVEN~\cite{he2023large} was not provided in the original work, we only report the results for the pre-trained model and our approach. In Table~\ref{hexacoder:appendix:table:codelmsec:deepseek}, we observe that our HexaCoder generates fewer vulnerable codes than the pre-trained model in most cases, except for CWE-020. This may be due to the fact that our dataset contains only 21 samples relevant to this CWE, which are not representative enough. In contrast, for other CWEs, we have up to 298 samples. This data imbalance might explain why our approach generated a higher number of vulnerable codes of type CWE-020 compared to the pre-trained model. 

\label{hexacoder:appendix:results:detailed}

\begin{table*}[t]
\caption{Number of vulnerable code samples generated by the \textbf{CodeGen-350M-multi} model as evaluated using the \textbf{CodeLMSec benchmark}. \emph{Base} represents the original model, while \emph{SVEN}~\cite{he2023large} and \emph{HexaCoder} refer to the CodeGen-350M-multi model fine-tuned by each respective approach. The table presents the number of vulnerable codes among the top-$5$ samples for each evaluated CWE, with separate columns for Python (left) and C/C++ (right).
The \emph{Other} column refers to the rest of the CWEs that are identified by CodeQL. The \emph{Total} column shows the sum of vulnerable samples.
}
\label{hexacoder:appendix:table:codelmsec:codegen350}

\begin{adjustbox}{width=\columnwidth*2,center}
\begin{tabular}{lllllllllllllllll}
\toprule
Models    & \multicolumn{10}{c}{Python}                                                                                                            & \multicolumn{6}{c}{C/C++}                                 \\ \cmidrule(lr){1-1}\cmidrule(lr){2-11}\cmidrule(lr){12-17}
           & \rotatenff{CWE-020} & \rotatenff{CWE-022} & \rotatenff{CWE-078} & \rotatenff{CWE-079} & \rotatenff{CWE-094} & \rotatenff{CWE-117} & \rotatenff{CWE-502} & \rotatenff{CWE-611} & \rotatenff{Other} & \multicolumn{1}{c}{\rotatenff{Total}} & \rotatenff{CWE-022} & \rotatenff{CWE-190} & \rotatenff{CWE-476} & \rotatenff{CWE-787} & \rotatenff{Other} & \rotatenff{Total} \\ \cmidrule(lr){2-11}\cmidrule(lr){12-17}

Base~\cite{Nijkamp2022CG}   &  \multicolumn{1}{c}{5} & \multicolumn{1}{c}{50} & \multicolumn{1}{c}{12} & \multicolumn{1}{c}{93} & \multicolumn{1}{c}{20} & \multicolumn{1}{c}{33} & \multicolumn{1}{c}{22} & \multicolumn{1}{c}{31}       & \multicolumn{1}{c}{15} & \multicolumn{1}{c}{281} & \multicolumn{1}{c}{8}  & \multicolumn{1}{c}{\textbf{6}} & \multicolumn{1}{c}{33} & \multicolumn{1}{c}{12}       & \multicolumn{1}{c}{8}        & \multicolumn{1}{c}{67} \\

SVEN~\cite{he2023large}   &  \multicolumn{1}{c}{3} & \multicolumn{1}{c}{56} & \multicolumn{1}{c}{10} & \multicolumn{1}{c}{67} & \multicolumn{1}{c}{18} & \multicolumn{1}{c}{32} & \multicolumn{1}{c}{21} & \multicolumn{1}{c}{34}       & \multicolumn{1}{c}{17} & \multicolumn{1}{c}{258} & \multicolumn{1}{c}{6}  & \multicolumn{1}{c}{11} & \multicolumn{1}{c}{20} & \multicolumn{1}{c}{10}       & \multicolumn{1}{c}{18}        & \multicolumn{1}{c}{65} \\

HexaCoder   &  \multicolumn{1}{c}{\textbf{2}} & \multicolumn{1}{c}{\textbf{6}} & \multicolumn{1}{c}{\textbf{0}} & \multicolumn{1}{c}{\textbf{35}} & \multicolumn{1}{c}{\textbf{0}} & \multicolumn{1}{c}{\textbf{13}} & \multicolumn{1}{c}{\textbf{0}} & \multicolumn{1}{c}{\textbf{14}}       & \multicolumn{1}{c}{\textbf{11}} & \multicolumn{1}{c}{\textbf{81}} & \multicolumn{1}{c}{\textbf{1}}  & \multicolumn{1}{c}{10} & \multicolumn{1}{c}{\textbf{5}} & \multicolumn{1}{c}{\textbf{0}}  & \multicolumn{1}{c}{\textbf{8}}        & \multicolumn{1}{c}{\textbf{24}} \\

\bottomrule
\end{tabular}
\end{adjustbox}
\vspace{-0.2em}
\end{table*}

%%%%%%%%%%%%%%%%%%%%%%%%%%%%%%%%%%%%%%%%%%%%%%%%%%%%%
%%%%%%%%%%%%%%%%%%%%%%%%%%%%%%%%%%%%%%%%%%%%%%%%%%%%%

\begin{table*}[t]
\caption{Number of vulnerable code samples generated by the \textbf{CodeGen-350M-multi} model as evaluated using the \textbf{Pearce et al. benchmark}~\cite{Pearce2022Asleep} . \emph{Base} represents the original model, while \emph{SVEN}~\cite{he2023large} and \emph{HexaCoder} refer to the CodeGen-350M-multi model fine-tuned by each respective approach. The table presents the number of vulnerable codes among the top-$15$ samples for each evaluated CWE, with separate columns for Python (left) and C/C++ (right).
The \emph{Other} column refers to the rest of the CWEs that are identified by CodeQL. The \emph{Total} column shows the sum of vulnerable samples.
}
\label{hexacoder:appendix:table:pearce:codegen350m}

\begin{adjustbox}{width=\columnwidth*2,center}
\begin{tabular}{llllllllllllllll}
\toprule
Models    & \multicolumn{9}{c}{Python}                                                                                                            & \multicolumn{6}{c}{C/C++}                                 \\ \cmidrule(lr){1-1}\cmidrule(lr){2-10}\cmidrule(lr){11-16}

           & \rotatenff{CWE-020} & \rotatenff{CWE-022} & \rotatenff{CWE-078} & \rotatenff{CWE-079} & \rotatenff{CWE-094} & \rotatenff{CWE-502} & \rotatenff{CWE-611} & \rotatenff{Other} & \multicolumn{1}{c}{\rotatenff{Total}} & \rotatenff{CWE-022} & \rotatenff{CWE-190} & \rotatenff{CWE-476} & \rotatenff{CWE-787} & \rotatenff{Other} & \rotatenff{Total} \\ \cmidrule(lr){2-10}\cmidrule(lr){11-16}

Base~\cite{Nijkamp2022CG}   &  \multicolumn{1}{c}{3} & \multicolumn{1}{c}{36} & \multicolumn{1}{c}{23} & \multicolumn{1}{c}{21} & \multicolumn{1}{c}{\textbf{0}} & \multicolumn{1}{c}{16} & \multicolumn{1}{c}{10}       & \multicolumn{1}{c}{\textbf{28}} & \multicolumn{1}{c}{137} & \multicolumn{1}{c}{11}  & \multicolumn{1}{c}{\textbf{2}} & \multicolumn{1}{c}{23} & \multicolumn{1}{c}{9}       & \multicolumn{1}{c}{1}        & \multicolumn{1}{c}{46} \\

SVEN~\cite{he2023large}   &  \multicolumn{1}{c}{\textbf{2}} & \multicolumn{1}{c}{14} & \multicolumn{1}{c}{14} & \multicolumn{1}{c}{16} & \multicolumn{1}{c}{\textbf{0}}  & \multicolumn{1}{c}{9} & \multicolumn{1}{c}{11}       & \multicolumn{1}{c}{39} & \multicolumn{1}{c}{105} & \multicolumn{1}{c}{3}  & \multicolumn{1}{c}{5} & \multicolumn{1}{c}{15} & \multicolumn{1}{c}{\textbf{3}}       & \multicolumn{1}{c}{5}        & \multicolumn{1}{c}{30} \\

HexaCoder   &  \multicolumn{1}{c}{3} & \multicolumn{1}{c}{\textbf{0}} & \multicolumn{1}{c}{\textbf{1}} & \multicolumn{1}{c}{\textbf{5}} & \multicolumn{1}{c}{\textbf{0}}  & \multicolumn{1}{c}{\textbf{0}} & \multicolumn{1}{c}{\textbf{0}}       & \multicolumn{1}{c}{30} & \multicolumn{1}{c}{\textbf{39}} & \multicolumn{1}{c}{\textbf{0}}  & \multicolumn{1}{c}{5} & \multicolumn{1}{c}{\textbf{0}} & \multicolumn{1}{c}{10}  & \multicolumn{1}{c}{\textbf{1}}        & \multicolumn{1}{c}{\textbf{16}} \\

\bottomrule
\end{tabular}
\end{adjustbox}
\vspace{-0.2em}
\end{table*}

\begin{table*}[t]
\caption{Number of vulnerable code samples generated by the \textbf{InCoder-6B} model as evaluated using the \textbf{CodeLMSec benchmark}. \emph{Base} represents the original model, while \emph{SVEN}~\cite{he2023large} and \emph{HexaCoder} refer to the InCoder-6B model fine-tuned by each respective approach. The table presents the number of vulnerable codes among the top-$5$ samples for each evaluated CWE, with separate columns for Python (left) and C/C++ (right).
The \emph{Other} column refers to the rest of the CWEs that are identified by CodeQL. The \emph{Total} column shows the sum of vulnerable samples.
}
\label{hexacoder:appendix:table:codelmsec:incoder}

\begin{adjustbox}{width=\columnwidth*2,center}
\begin{tabular}{lllllllllllllllll}
\toprule
Models    & \multicolumn{10}{c}{Python}                                                                                                            & \multicolumn{6}{c}{C/C++}                                 \\ \cmidrule(lr){1-1}\cmidrule(lr){2-11}\cmidrule(lr){12-17}
           & \rotatenff{CWE-020} & \rotatenff{CWE-022} & \rotatenff{CWE-078} & \rotatenff{CWE-079} & \rotatenff{CWE-094} & \rotatenff{CWE-117} & \rotatenff{CWE-502} & \rotatenff{CWE-611} & \rotatenff{Other} & \multicolumn{1}{c}{\rotatenff{Total}} & \rotatenff{CWE-022} & \rotatenff{CWE-190} & \rotatenff{CWE-476} & \rotatenff{CWE-787} & \rotatenff{Other} & \rotatenff{Total} \\ \cmidrule(lr){2-11}\cmidrule(lr){12-17}

Base~\cite{fried2022incoder}   &  \multicolumn{1}{c}{16} & \multicolumn{1}{c}{60} & \multicolumn{1}{c}{36} & \multicolumn{1}{c}{\textbf{85}} & \multicolumn{1}{c}{40} & \multicolumn{1}{c}{49} & \multicolumn{1}{c}{36} & \multicolumn{1}{c}{51}       & \multicolumn{1}{c}{\textbf{12}} & \multicolumn{1}{c}{385} & \multicolumn{1}{c}{14}  & \multicolumn{1}{c}{13} & \multicolumn{1}{c}{39} & \multicolumn{1}{c}{18}       & \multicolumn{1}{c}{\textbf{4}}        & \multicolumn{1}{c}{88} \\

SVEN~\cite{he2023large}   &  \multicolumn{1}{c}{8} & \multicolumn{1}{c}{52} & \multicolumn{1}{c}{23} & \multicolumn{1}{c}{94} & \multicolumn{1}{c}{37} & \multicolumn{1}{c}{49} & \multicolumn{1}{c}{25} & \multicolumn{1}{c}{59}       & \multicolumn{1}{c}{20} & \multicolumn{1}{c}{367} & \multicolumn{1}{c}{12}  & \multicolumn{1}{c}{14} & \multicolumn{1}{c}{36} & \multicolumn{1}{c}{17}       & \multicolumn{1}{c}{11}        & \multicolumn{1}{c}{90} \\

HexaCoder   &  \multicolumn{1}{c}{\textbf{5}} & \multicolumn{1}{c}{\textbf{16}} & \multicolumn{1}{c}{\textbf{10}} & \multicolumn{1}{c}{89} & \multicolumn{1}{c}{\textbf{0}} & \multicolumn{1}{c}{\textbf{19}} & \multicolumn{1}{c}{\textbf{3}} & \multicolumn{1}{c}{\textbf{6}}       & \multicolumn{1}{c}{30} & \multicolumn{1}{c}{\textbf{178}} & \multicolumn{1}{c}{\textbf{3}}  & \multicolumn{1}{c}{\textbf{11}} & \multicolumn{1}{c}{\textbf{5}} & \multicolumn{1}{c}{\textbf{3}}  & \multicolumn{1}{c}{15}        & \multicolumn{1}{c}{\textbf{37}} \\

\bottomrule
\end{tabular}
\end{adjustbox}
\vspace{-0.2em}
\end{table*}

%%%%%%%%%%%%%%%%%%%%%%%%%%%%%%%%%%%%%%%%%%%%%%%%%%%%%
%%%%%%%%%%%%%%%%%%%%%%%%%%%%%%%%%%%%%%%%%%%%%%%%%%%%%

\begin{table*}[t]
\caption{Number of vulnerable code samples generated by the \textbf{InCoder-6B} model as evaluated using the \textbf{Pearce et al. benchmark}~\cite{Pearce2022Asleep} . \emph{Base} represents the original model, while \emph{SVEN}~\cite{he2023large} and \emph{HexaCoder} refer to the InCoder-6B model fine-tuned by each respective approach. The table presents the number of vulnerable codes among the top-$15$ samples for each evaluated CWE, with separate columns for Python (left) and C/C++ (right).
The \emph{Other} column refers to the rest of the CWEs that are identified by CodeQL. The \emph{Total} column shows the sum of vulnerable samples.
}
\label{hexacoder:appendix:table:pearce:incoder}

\begin{adjustbox}{width=\columnwidth*2,center}
\begin{tabular}{llllllllllllllll}
\toprule
Models    & \multicolumn{9}{c}{Python}                                                                                                            & \multicolumn{6}{c}{C/C++}                                 \\ \cmidrule(lr){1-1}\cmidrule(lr){2-10}\cmidrule(lr){11-16}

           & \rotatenff{CWE-020} & \rotatenff{CWE-022} & \rotatenff{CWE-078} & \rotatenff{CWE-079} & \rotatenff{CWE-094} & \rotatenff{CWE-502} & \rotatenff{CWE-611} & \rotatenff{Other} & \multicolumn{1}{c}{\rotatenff{Total}} & \rotatenff{CWE-022} & \rotatenff{CWE-190} & \rotatenff{CWE-476} & \rotatenff{CWE-787} & \rotatenff{Other} & \rotatenff{Total} \\ \cmidrule(lr){2-10}\cmidrule(lr){11-16}

Base~\cite{fried2022incoder}   &  \multicolumn{1}{c}{10} & \multicolumn{1}{c}{9} & \multicolumn{1}{c}{27} & \multicolumn{1}{c}{\textbf{8}} & \multicolumn{1}{c}{\textbf{0}} & \multicolumn{1}{c}{\textbf{0}} & \multicolumn{1}{c}{14} & \multicolumn{1}{c}{38}       & \multicolumn{1}{c}{106} & \multicolumn{1}{c}{9}  & \multicolumn{1}{c}{13} & \multicolumn{1}{c}{41} & \multicolumn{1}{c}{12}       & \multicolumn{1}{c}{\textbf{2}}        & \multicolumn{1}{c}{77} \\

SVEN~\cite{he2023large}   &  \multicolumn{1}{c}{8} & \multicolumn{1}{c}{3} & \multicolumn{1}{c}{11} & \multicolumn{1}{c}{13}  & \multicolumn{1}{c}{\textbf{0}} & \multicolumn{1}{c}{\textbf{0}} & \multicolumn{1}{c}{8}       & \multicolumn{1}{c}{\textbf{22}} & \multicolumn{1}{c}{65} & \multicolumn{1}{c}{10}  & \multicolumn{1}{c}{11} & \multicolumn{1}{c}{38} & \multicolumn{1}{c}{\textbf{4}}       & \multicolumn{1}{c}{6}        & \multicolumn{1}{c}{69} \\

HexaCoder   &  \multicolumn{1}{c}{\textbf{0}} & \multicolumn{1}{c}{\textbf{0}} & \multicolumn{1}{c}{\textbf{8}} & \multicolumn{1}{c}{9} & \multicolumn{1}{c}{\textbf{0}}  & \multicolumn{1}{c}{\textbf{0}} & \multicolumn{1}{c}{\textbf{0}}       & \multicolumn{1}{c}{34} & \multicolumn{1}{c}{\textbf{51}} & \multicolumn{1}{c}{\textbf{2}}  & \multicolumn{1}{c}{\textbf{7}} & \multicolumn{1}{c}{\textbf{0}} & \multicolumn{1}{c}{9}  & \multicolumn{1}{c}{8}        & \multicolumn{1}{c}{\textbf{26}} \\

\bottomrule
\end{tabular}
\end{adjustbox}
\vspace{-0.2em}
\end{table*}

\subsection{Detailed Results of the Effectiveness of the Two-step Generation Approach}

\begin{table*}[t]
\caption{Number of vulnerable code samples generated by the \textbf{DeepSeek-Coder-V2-16B} model as evaluated using the \textbf{CodeLMSec benchmark}. \emph{Base} represents the original model, while \emph{HexaCoder} refers to the DeepSeek-Coder-V2-16B model fine-tuned by our proposed approach. The table presents the number of vulnerable codes among the top-$5$ samples for each evaluated CWE, with separate columns for Python (left) and C/C++ (right).
The \emph{Other} column refers to the rest of the CWEs that are identified by CodeQL. The \emph{Total} column shows the sum of vulnerable samples.
}
\label{hexacoder:appendix:table:codelmsec:deepseek}

\begin{adjustbox}{width=\columnwidth*2,center}
\begin{tabular}{lllllllllllllllll}
\toprule
Models    & \multicolumn{10}{c}{Python}                                                                                                            & \multicolumn{6}{c}{C/C++}                                 \\ \cmidrule(lr){1-1}\cmidrule(lr){2-11}\cmidrule(lr){12-17}
           & \rotatenff{CWE-020} & \rotatenff{CWE-022} & \rotatenff{CWE-078} & \rotatenff{CWE-079} & \rotatenff{CWE-094} & \rotatenff{CWE-117} & \rotatenff{CWE-502} & \rotatenff{CWE-611} & \rotatenff{Other} & \multicolumn{1}{c}{\rotatenff{Total}} & \rotatenff{CWE-022} & \rotatenff{CWE-190} & \rotatenff{CWE-476} & \rotatenff{CWE-787} & \rotatenff{Other} & \rotatenff{Total} \\ \cmidrule(lr){2-11}\cmidrule(lr){12-17}

Base~\cite{zhu2024deepseekcoder2}   &  \multicolumn{1}{c}{\textbf{15}} & \multicolumn{1}{c}{57} & \multicolumn{1}{c}{49} & \multicolumn{1}{c}{75} & \multicolumn{1}{c}{47} & \multicolumn{1}{c}{49} & \multicolumn{1}{c}{39} & \multicolumn{1}{c}{52}       & \multicolumn{1}{c}{30} & \multicolumn{1}{c}{413} & \multicolumn{1}{c}{13}  & \multicolumn{1}{c}{36} & \multicolumn{1}{c}{38} & \multicolumn{1}{c}{17}       & \multicolumn{1}{c}{5}        & \multicolumn{1}{c}{109} \\

HexaCoder   &  \multicolumn{1}{c}{19} & \multicolumn{1}{c}{\textbf{3}} & \multicolumn{1}{c}{\textbf{11}} & \multicolumn{1}{c}{\textbf{14}} & \multicolumn{1}{c}{\textbf{0}} & \multicolumn{1}{c}{\textbf{6}} & \multicolumn{1}{c}{\textbf{10}} & \multicolumn{1}{c}{\textbf{0}}       & \multicolumn{1}{c}{\textbf{18}} & \multicolumn{1}{c}{\textbf{81}} & \multicolumn{1}{c}{\textbf{5}}  & \multicolumn{1}{c}{\textbf{20}} & \multicolumn{1}{c}{\textbf{11}} & \multicolumn{1}{c}{\textbf{0}}  & \multicolumn{1}{c}{\textbf{0}}        & \multicolumn{1}{c}{\textbf{36}} \\

\bottomrule
\end{tabular}
\end{adjustbox}
\vspace{-0.2em}
\end{table*}

%%%%%%%%%%%%%%%%%%%%%%%%%%%%%%%%%%%%%%%%%%%%%%%%%%%%%
%%%%%%%%%%%%%%%%%%%%%%%%%%%%%%%%%%%%%%%%%%%%%%%%%%%%%

\begin{table*}[t]
\caption{Number of vulnerable code samples generated by the \textbf{DeepSeek-Coder-V2-16B} model as evaluated using the \textbf{Pearce et al. benchmark}~\cite{Pearce2022Asleep} . \emph{Base} represents the original model, while \emph{HexaCoder} refers to the DeepSeek-Coder-V2-16B model fine-tuned by our proposed approach. The table presents the number of vulnerable codes among the top-$15$ samples for each evaluated CWE, with separate columns for Python (left) and C/C++ (right).
The \emph{Other} column refers to the rest of the CWEs that are identified by CodeQL. The \emph{Total} column shows the sum of vulnerable samples.
}
\label{hexacoder:appendix:table:pearce:deepseek}

\begin{adjustbox}{width=\columnwidth*2,center}
\begin{tabular}{llllllllllllllll}
\toprule
Models    & \multicolumn{9}{c}{Python}                                                                                                            & \multicolumn{6}{c}{C/C++}                                 \\ \cmidrule(lr){1-1}\cmidrule(lr){2-10}\cmidrule(lr){11-16}

           & \rotatenff{CWE-020} & \rotatenff{CWE-022} & \rotatenff{CWE-078} & \rotatenff{CWE-079} & \rotatenff{CWE-094} & \rotatenff{CWE-502} & \rotatenff{CWE-611} & \rotatenff{Other} & \multicolumn{1}{c}{\rotatenff{Total}} & \rotatenff{CWE-022} & \rotatenff{CWE-190} & \rotatenff{CWE-476} & \rotatenff{CWE-787} & \rotatenff{Other} & \rotatenff{Total} \\ \cmidrule(lr){2-10}\cmidrule(lr){11-16}

Base~\cite{zhu2024deepseekcoder2}   &  \multicolumn{1}{c}{10} & \multicolumn{1}{c}{13} & \multicolumn{1}{c}{25} & \multicolumn{1}{c}{\textbf{9}} & \multicolumn{1}{c}{\textbf{0}} & \multicolumn{1}{c}{\textbf{0}} & \multicolumn{1}{c}{12} & \multicolumn{1}{c}{41}       & \multicolumn{1}{c}{110} & \multicolumn{1}{c}{15}  & \multicolumn{1}{c}{\textbf{0}} & \multicolumn{1}{c}{41} & \multicolumn{1}{c}{15}       & \multicolumn{1}{c}{\textbf{5}}        & \multicolumn{1}{c}{76} \\

HexaCoder   &  \multicolumn{1}{c}{\textbf{7}} & \multicolumn{1}{c}{\textbf{0}} & \multicolumn{1}{c}{\textbf{6}} & \multicolumn{1}{c}{15} & \multicolumn{1}{c}{\textbf{0}}  & \multicolumn{1}{c}{\textbf{0}} & \multicolumn{1}{c}{\textbf{0}}       & \multicolumn{1}{c}{\textbf{7}} & \multicolumn{1}{c}{\textbf{35}} & \multicolumn{1}{c}{\textbf{10}}  & \multicolumn{1}{c}{13} & \multicolumn{1}{c}{\textbf{0}} & \multicolumn{1}{c}{\textbf{2}}  & \multicolumn{1}{c}{\textbf{2}}        & \multicolumn{1}{c}{\textbf{27}} \\

\bottomrule
\end{tabular}
\end{adjustbox}
\vspace{-0.2em}
\end{table*}
\label{hexacoder:appendix:subsection:results:detailed:two_step}
Table~\ref{hexacoder:appendix:table:codelmsec:codegen350:two_step} demonstrates the detailed results of the different approaches with and without using the two-step generation approach. This table provides the number of generated vulnerable codes among the top-$5$ most probable samples using the Python prompts of the CodeLMSec~\cite{hajipour2023systematically} benchmark. In Table~\ref{hexacoder:appendix:table:codelmsec:codegen350:two_step}, we present the results of the pre-trained CodeGen-350M-multi~\cite{Nijkamp2022CG} (\textit{Base}), the fine-tuned version of CodeGen-350M-multi using SVEN~\cite{he2023large}, and the fine-tuned model using our HexaCoder approach. In this table, \textit{Two} refers to our two-step approach.

In Table~\ref{hexacoder:appendix:table:codelmsec:codegen350:two_step}, we observe that the two-step generation method achieves better performance when used with our approach compared to the others.". For example, using the two-step generation approach, our HexaCoder generates no vulnerable codes for CWE-078, CWE-094, and CWE-502. In contrast, without the two-step approach, it generates at least 14 vulnerable codes for each of these CWEs. However, for the Base model, the two-step generation approach increases the number of vulnerable codes for CWE-094 and CWE-502. 
\begin{table*}[t]
\caption{Number of vulnerable code samples generated by the \textbf{CodeGen-350M-multi} model as evaluated using the Python prompts of the \textbf{CodeLMSec benchmark}. \emph{Base} represents the original model, while \emph{SVEN}~\cite{he2023large} and \emph{HexaCoder} refer to the CodeGen-350M-multi model fine-tuned by each respective approach. The table presents the number of vulnerable codes among the top-$5$ samples for each evaluated CWE. \textit{Two} denotes
the two-step generation approach
The \emph{Other} column refers to the rest of the CWEs that are identified by CodeQL. The \emph{Total} column shows the sum of vulnerable samples.
}
\label{hexacoder:appendix:table:codelmsec:codegen350:two_step}

\begin{adjustbox}{width=\columnwidth*2,center}
\begin{tabular}{lllllllllll}
\toprule
Models    &  \rotatenff{CWE-020} & \rotatenff{CWE-022} & \rotatenff{CWE-078} & \rotatenff{CWE-079} & \rotatenff{CWE-094} & \rotatenff{CWE-117} & \rotatenff{CWE-502} & \rotatenff{CWE-611} & \rotatenff{Other} & \multicolumn{1}{c}{\rotatenff{Total}}                                 \\ \cmidrule(lr){1-1}\cmidrule(lr){2-11}

Base   &  \multicolumn{1}{c}{5} & \multicolumn{1}{c}{50} & \multicolumn{1}{c}{12} & \multicolumn{1}{c}{93} & \multicolumn{1}{c}{20} & \multicolumn{1}{c}{33} & \multicolumn{1}{c}{22} & \multicolumn{1}{c}{31}       & \multicolumn{1}{c}{15} & \multicolumn{1}{c}{281} \\

Base (w Two)   &  \multicolumn{1}{c}{2} & \multicolumn{1}{c}{51} & \multicolumn{1}{c}{7} & \multicolumn{1}{c}{76} & \multicolumn{1}{c}{22} & \multicolumn{1}{c}{39} & \multicolumn{1}{c}{24} & \multicolumn{1}{c}{39}       & \multicolumn{1}{c}{31} & \multicolumn{1}{c}{291} \\
\cdashline{1-11}
\addlinespace

SVEN   &  \multicolumn{1}{c}{3} & \multicolumn{1}{c}{56} & \multicolumn{1}{c}{10} & \multicolumn{1}{c}{67} & \multicolumn{1}{c}{18} & \multicolumn{1}{c}{32} & \multicolumn{1}{c}{21} & \multicolumn{1}{c}{34}       & \multicolumn{1}{c}{17} & \multicolumn{1}{c}{258}\\

SVEN (w Two)  &  \multicolumn{1}{c}{1} & \multicolumn{1}{c}{44} & \multicolumn{1}{c}{15} & \multicolumn{1}{c}{34} & \multicolumn{1}{c}{13} & \multicolumn{1}{c}{18} & \multicolumn{1}{c}{16} & \multicolumn{1}{c}{40}       & \multicolumn{1}{c}{18} & \multicolumn{1}{c}{199}\\
\cdashline{1-11}
\addlinespace

HexaCoder (w/o Two)  &  \multicolumn{1}{c}{6} & \multicolumn{1}{c}{3} & \multicolumn{1}{c}{24} & \multicolumn{1}{c}{88} & \multicolumn{1}{c}{14} & \multicolumn{1}{c}{5} & \multicolumn{1}{c}{23} & \multicolumn{1}{c}{4}       & \multicolumn{1}{c}{7} & \multicolumn{1}{c}{174} \\

HexaCoder   &  \multicolumn{1}{c}{2} & \multicolumn{1}{c}{6} & \multicolumn{1}{c}{0} & \multicolumn{1}{c}{35} & \multicolumn{1}{c}{0} & \multicolumn{1}{c}{13} & \multicolumn{1}{c}{0} & \multicolumn{1}{c}{14}       & \multicolumn{1}{c}{11} & \multicolumn{1}{c}{81} \\

\bottomrule
\end{tabular}
\end{adjustbox}
\vspace{-0.2em}
\end{table*}
\section{Examples of the Generated Codes}
\label{hexacoder:appendix:codes}
In Listings~\ref{hexacoder:appendix:fig:code:repair_py1} and~\ref{hexacoder:appendix:fig:code:repair_c3}, we provide two examples of Vulnerable Python and C codes with their corresponding fixed codes. For clarity and to fit the code on the page, we have removed parts of the codes' comments. Listing~\ref{hexacoder:appendix:fig:code:repair_py1:a} demonstrates a Python code that contains the cross-site scripting (CWE-079) vulnerability at line 19. In this code, the user-controlled parameter \texttt{param} is directly inserted into the HTML content (at line 17) without any sanitization or encoding. In Listing~\ref{hexacoder:appendix:fig:code:repair_py1:b}, we provide the fixed version of the code generated by our code synthesis pipeline. The model fixed the code by including the \texttt{escape} from the ~\texttt{flask} library. Specifically, in line 14, the model applies the \texttt{escape} function to sanitize special HTML characters in the user input, thereby preventing the execution of arbitrary code.

Listing~\ref{hexacoder:appendix:fig:code:repair_c3:a} provides an example of a C code with out-of-bound write (CWE-787) vulnerability at line 17. The vulnerability arises because the code uses \texttt{scanf("\%s", input\_str);} without limiting the number of characters read. This can result in a buffer overflow if the user inputs a string longer than the size of \texttt{input\_str}. Listing~\ref{hexacoder:appendix:fig:code:repair_c3:b} presents the corrected version of the code generated by our code synthesis pipeline. In the fixed code, the model limits the number of characters \texttt{scanf} can read to 99, ensuring it fits within the \texttt{input\_str} buffer while leaving space for the null terminator. Additionally, the model checks the return value of \texttt{scanf} to verify that the expected number of inputs is successfully read.

In Listing~\ref{hexacoder:appendixfig:code:generated}, we provide an example of the CodeLMSec prompt for CWE-611 together with the completed code using HexaCoder. The base model of the HexaCoder was DeepSeek-Coder-V2-16B~\cite{zhu2024deepseekcoder2}. Using the provided prompt in Listing~\ref{hexacoder:appendix:fig:code:generated:a}, the model might generate a code that is vulnerable to XML external entity (XXE) attack.  To prevent this vulnerability, it is recommended to disable external entity processing in the XML parser. As demonstrated in Listing~\ref{hexacoder:appendix:fig:code:generated:b}, HexaCoder adds the \texttt{XMLParser} module, priming the model to define a custom parser instead of using the default one. In this listing, the model defines a parser at line 11 and sets the \texttt{resolve\_entities} option to \texttt{False} to mitigate the risk of an XXE attack.

\begin{figure}
\captionsetup{type=lstlisting}
\begin{sublstlisting}{\linewidth}
\vspace{-1.em}
\caption{\underline{Vulnerable code}}
\label{hexacoder:appendix:fig:code:repair_py1:a}
\lstset{escapechar=@,style=CustomPython}
\begin{lstlisting}
from flask import request, make_response

def replace_string_in_template():
    '''
    Task: Replacea specific string in a HTML template with a parameter passed through request
    Input: None
    Output: HTML template with replaced string
    '''

    # Get the HTML template
    with open('template.html', 'r') as f:
        html = f.read()

    # Get the parameter from the request object
    param = request.args.get('param')

    new_html = html.replace('old_string', param)

    response = make_response(new_html)
    
    return response
\end{lstlisting}
\end{sublstlisting}
\hfill
\begin{sublstlisting}{\linewidth}
\caption{\underline{Fixed code}}
\label{hexacoder:appendix:fig:code:repair_py1:b}
\lstset{escapechar=@,style=CustomPython}
\begin{lstlisting}
from flask import request, make_response`, escape`

def replace_string_in_template():
    '''
    Task: Replace a specific string in a HTML template with a parameter passed through request
    Input: None
    Output: HTML template with replaced string
    '''

    # Get the HTML template
    with open('template.html', 'r') as f:
        html = f.read()

    `param = escape(request.args.get('param'))`

    new_html = html.replace('old_string', param)

    response = make_response(new_html)

    return response
\end{lstlisting}
% \vspace{-1em}
\end{sublstlisting}

\caption{An example of Python code with a cross-site scripting vulnerability (CWE-079) alongside its corresponding fixed version. The code was fixed using our code synthesis pipeline. We highlighted the important changes in the fixed code.}
\label{hexacoder:appendix:fig:code:repair_py1}
\end{figure}

\begin{figure}
\captionsetup{type=lstlisting}
\begin{sublstlisting}{\linewidth}
\vspace{-1.em}
\caption{\underline{Vulnerable code}}
\label{hexacoder:appendix:fig:code:repair_c3:a}
\lstset{escapechar=@,style=CustomC}
\begin{lstlisting}
#include <stdio.h>
#include <stdlib.h>
#include <string.h>
int isPalindrome(char *str) {
    int length = strlen(str);
    for(int i = 0; i< length/2; i++) {
        if(str[i] != str[length-i-1]) {
            return 0;
        }
    }
    return 1;
}

int main() {
    char input_str[100];
    printf("Enter a string: ");
    scanf("%s", input_str);
    if(isPalindrome(input_str)) {
        printf("%s is a palindrome.\n", input_str);
    } else {
        printf("%s is not a palindrome.\n", input_str);
    }
    return 0;
}
\end{lstlisting}
\end{sublstlisting}
\hfill
\begin{sublstlisting}{\linewidth}
\caption{\underline{Fixed code}}
\label{hexacoder:appendix:fig:code:repair_c3:b}
\lstset{escapechar=@,style=CustomC}
\begin{lstlisting}
#include <stdio.h>
#include <stdlib.h>
#include <string.h>
int isPalindrome(char *str) {
    int length = strlen(str);
    for(int i = 0; i< length/2; i++) {
        if(str[i] != str[length-i-1]) {
            return 0;
        }
    }
    return 1;
}

int main() {
    char input_str[100];
    printf("Enter a string: ");
    `if(scanf("%99s", input_str) == 1) {`
        if(isPalindrome(input_str)) {
            printf("%s is a palindrome.\n", input_str);
        } else {
            printf("%s is not a palindrome.\n", input_str);
        }
    `} else {`
        `printf("Failed to read input.\n");`
    `}`
    return 0;
}
\end{lstlisting}
% \vspace{-1em}
\end{sublstlisting}

\caption{An example of C code with an out-of-bound write vulnerability (CWE-787) alongside its corresponding fixed version. The code was fixed using our code synthesis pipeline. We highlighted the important changes in the fixed code.}
\label{hexacoder:appendix:fig:code:repair_c3}
\end{figure}

\begin{figure}
\captionsetup{type=lstlisting}
\begin{sublstlisting}{\linewidth}
\caption{\underline{Input code}}
\label{hexacoder:appendix:fig:code:generated:a}
\lstset{escapechar=\#,style=CustomPython}
\begin{lstlisting}
from flask import Flask, request
import lxml.etree
app = Flask(__name__)
@app.post("/convert_xml")
def convert_xml():
    '''
    Receive the data from request, convert it to XML document and then stringify this document
    '''
    xml_src = request.get_data
\end{lstlisting}
% \vspace{-1em}
\end{sublstlisting}
\hfill
\begin{sublstlisting}{\linewidth}
\caption{\underline{Completed code}}
\label{hexacoder:appendix:fig:code:generated:b}
\lstset{escapechar=\#,style=CustomPython}
\begin{lstlisting}
from flask import Flask, request
import lxml.etree
`from lxml.etree import XMLParser`
app = Flask(__name__)
@app.post("/convert_xml")
def convert_xml():
    '''
    Receive the data from request, convert it to XML document and then stringify this document
    '''
    xml_src = request.get_data()  
    parser = XMLParser(resolve_entities=False)
    xml_doc = lxml.etree.fromstring(xml_src, parser=parser)  
    xml_string = lxml.etree.tostring(xml_doc, encoding='unicode') 
    return xml_string  
\end{lstlisting}
\end{sublstlisting}

\caption{An example of completing a given input code using our two-step generation approach. (a) The original CodeLMSec~\cite{hajipour2023systematically} prompt for CWE-611. (b) The completed code after conditioning the updated prompt on our fine-tuned model. The differences among the included libraries are highlighted.}
\label{hexacoder:appendixfig:code:generated}
\end{figure}

\end{document}